\documentclass[useAMS]{mn2e}
\usepackage{times}
\input{epsf}
\usepackage{graphicx}
\usepackage{lscape}
\usepackage{psfig}
\usepackage{rotating}
\usepackage{float}

\newcommand{\xmmn}{{\it XMM-Newton~\/}}

\newcommand{\chan}{{\it Chandra~\/}}

\newcommand{\sas}{{\sc{sas~}}}

\newcommand{\ciao}{{\sc{ciao~}}}
\newcommand{\xspec}{{\sc{xspec~}}}
\newcommand{\chart}{{\sc{chart~}}}

\newcommand{\vla}{{\it VLA~\/}}

\newcommand{\colden}{{\sc{colden~}}}

\newcommand{\nvss}{{\sc{NVSS\/}}}
\newcommand{\first}{{\sc{FIRST\/}}}

\def\lx{L$_{\rm X}$}
\def\sfrx{SFR$_{\rm X}$}

\def\la{\mathrel{\hbox{\rlap{\hbox{\lower4pt\hbox{$\sim$}}}{\raise2pt\hbox{$<$}}}}}
\def\ga{\mathrel{\hbox{\rlap{\hbox{\lower4pt\hbox{$\sim$}}}{\raise2pt\hbox{$>$}}}}}

\title[High X-ray luminosities in SDSS galaxies]
{On the nature of high X-ray luminosities in SDSS galaxies}

\author[F.\,E. Jackson et al.]
{F.\,E. Jackson$^{1}$\thanks{E-mail: f.e.jackson@durham.ac.uk}, T.\,P. Roberts$^{1}$, D.\,M. Alexander$^{1}$, J.\,M. Gelbord$^{1,2}$, A.\,D. Goulding$^{1,3}$, \and M.\,J. Ward$^{1}$, J.\,L. Wardlow$^{1,4}$, M.\,G. Watson$^{5}$\\
$^{1}$Department of Physics, University of Durham, South Road, Durham DH1 3LE, UK\\ $^2$ Astronomy \& Astrophysics, Eberly College of Science, The Pennsylvania State University, 525 Davey Lab, University Park, PA 16802, USA\\ $^3$Harvard-Smithsonian Center for Astrophysics, 60 Garden Street, Cambridge, MA 02138, USA \\$^4$ Department of Physics \& Astronomy, 4129 Fredrick Reines Hall, University of California, Irvine, CA 92697-4575, USA\\ $^{5}$X-ray \& Observational Astronomy Group, Dept. of Physics \& Astronomy, University of Leicester, University Road, Leicester LE1 7RH, UK}

\date{Submitted to MNRAS}
\pagerange{\pageref{firstpage}--\pageref{lastpage}}
\pubyear{2011}

\begin{document}
\label{firstpage}
\maketitle

\begin{abstract}
%\begin{center}
%\hspace{-20mm}

Surveys have revealed a class of object displaying both high X-ray
luminosities (L$_{\rm X}$ $>$ 10$^{42}$ erg s$^{-1}$), and a lack of a
discernible active galactic nucleus (AGN) in the optical band. If
these sources are powered by star formation activity alone, they would
be the most extreme X-ray luminosity star forming galaxies known. We
have investigated the mechanism driving the X-ray luminosities of
such galaxies by studying the X-ray emission of three moderate redshift (z
$\sim$ 0.1) examples of this class, selected from a cross-correlation
of the SDSS-DR5 and 2XMMp-DR0 catalogues. X-ray spatial and long-term
variability diagnostics of these sources suggest that they are compact
X-ray emitters. This result is supported by the detection of rapid
short term variability in an observation of one of the sources. The X-ray
spectra of all three sources are best fitted with a simple absorbed
power-law model, thus betraying no significant signs of star
formation. These results indicate that the X-ray emission is powered
by AGN activity. But why do these sources not display
optical AGN signatures? We show that the most likely explanation is
that the optical AGN emission lines are being diluted by star
formation signatures from within their host galaxies.

\end{abstract}

	
\section{Introduction}

\begin{figure}
\begin{center}
\hspace{-6mm}
\vbox{
\psfig{figure=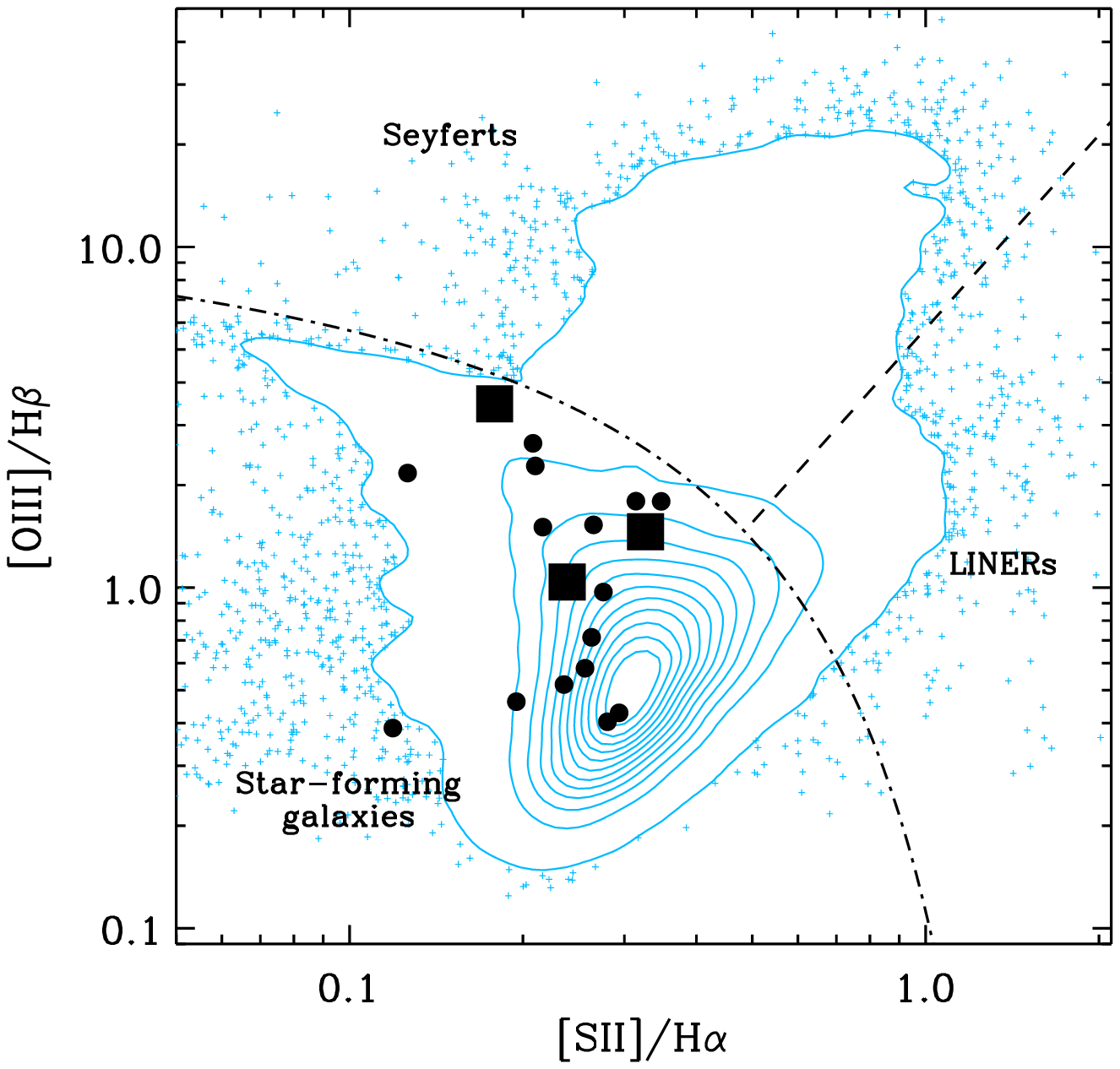,height=8.4cm}
\psfig{figure=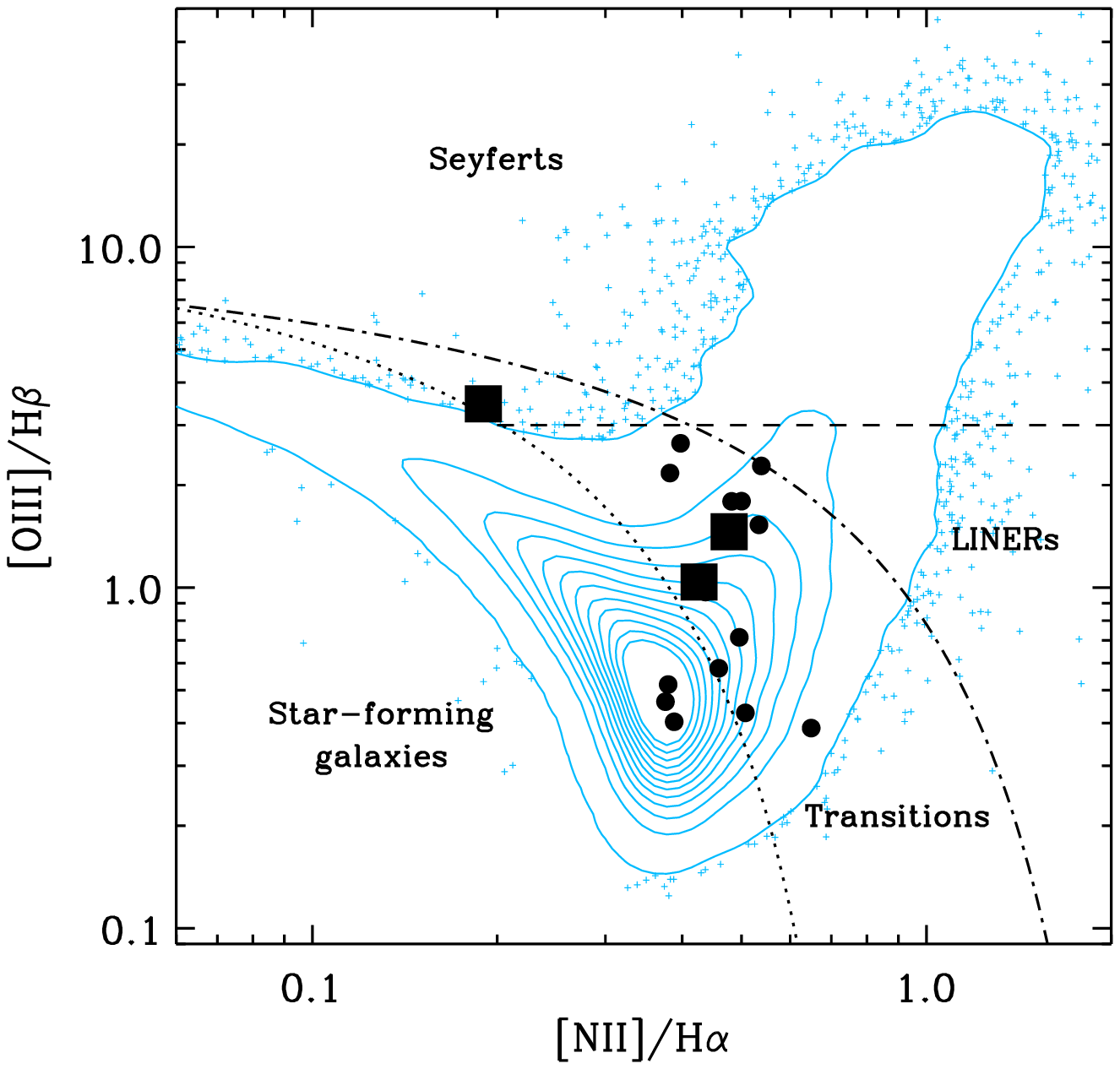,height=8.4cm}
} 
\vspace{-5mm}
\caption {Emission-line diagnostic diagrams displaying the [S{\small II}]/H$\alpha$ vs. [O{\small III}]/H$\beta$ ({\it top panel}) and [N{\small II}]/H$\alpha$ vs. [O{\small III}]/H$\beta$ ({\it bottom panel}) line ratios of a large sample of galaxies, selected from the SDSS catalogue. Among these galaxies are the three sources being studied in this work, whose flux ratios are represented by the filled squares. The filled circles represent the flux ratios of the 16 remaining sources from the cross-correlation of SDSS and 2XMMp that were selected for their high X-ray luminosity (i.e., a luminosity $>$ 10$^{42}$ erg s$^{-1}$). In both panels, the dot-dash curves represent the Kewley et al. (2001) theoretical maximum starburst line. The dotted curve in the bottom panel represents the Kauffman et al. (2003) border between the star forming region and the LINER/Seyfert regions of the diagram. The dashed lines in both panels represent the Kauffman et al. (2003) border between the diagrams' LINER and Seyfert regions. The blue contours represent the density of line ratio measurements from a
large sample of galaxies, selected from the SDSS DR7 catalogue, with the
blue crosses representing individual galaxies in the outlying region
beyond the lowest contour.}
%\caption {Emission-line diagnostic diagrams displaying the [S{\small II}]/H$\alpha$ versus [O{\small III}]/H$\beta$ line ratios of a large sample of galaxies, selected from the SDSS catalogue. Among these galaxies are the three sources being studied in this work, whose flux ratios are represented by the filled squares. The filled circles  represent the flux ratios of the 16 remaining sources from the cross-correlation of SDSS and 2XMMp that were selected for their high X-ray luminosity (i.e., a luminosity $>$ 10$^{42}$ erg s$^{-1}$). The dot-dash curves represent the Kewley et al. (2001) theoretical maximum starburst line. The dashed line represents the Kauffman et al. (2003) border between the diagrams' LINER and Seyfert regions.}

\label{fig:bpt}
\end{center}
\end{figure}

Deep and wide area X-ray surveys (e.g.,\ Alexander et~al. 2003; Lehmer
et~al. 2005; Kim et~al. 2007; Elvis et~al. 2009; Watson et~al. 2009;
Xue et~al. 2010) have resolved up to $\approx$~90\% of the X-ray
background into discrete sources over the 0.5--8~keV band (e.g.,\ Lumb
et~al. 2002; Moretti et~al. 2003; Bauer et~al. 2004; Worsley
et~al. 2005; Hickox \& Markevitch 2006). The majority of the objects
detected in these surveys are luminous X-ray sources at
$z\approx$~0.2--5 with $L_{\rm X}\geq10^{42}$~erg~s$^{-1}$ (e.g.,\
Hornschemeier et~al. 2001; Barger 2003; Brandt \& Hasinger
2005; Treister et~al. 2005; Silverman et~al. 2010). Many of these
luminous X-ray sources are clearly Active Galactic Nuclei (AGN), as
demonstrated by the detection of either optical AGN signatures (i.e.,\
broad emission lines and/or narrow high-excitation emission lines;
e.g.,\ Barger et~al. 2002; Szokoly et~al. 2004) or hard X-ray spectra
($\Gamma\leq1$; Mushotzky et~al. 2000; Giacconi
et~al. 2001). However, a significant fraction of these luminous X-ray
sources display %star formation-like X-ray emission (i.e., display an optically thin, soft ($\Gamma > 1$), thermal spectrum, likely having multiple temperature components in the 0.1 - 1 keV range, e.g., Moran et al. 1999; Jenkins et al. 2004; Warwick et al. 2007), and 
optical spectroscopic signatures consistent with star formation (e.g.,\ Barger
et~al. 2002; Szokoly et~al. 2004; Eckart et~al. 2006; Trouille
et~al. 2008). %This 
The luminous X-ray emission could be due to either AGN activity where the
optical spectroscopic signatures are extinguished or diluted (e.g.,\
Moran et~al. 2002; Maiolino et~al. 2003; Goulding \& Alexander 2009),
or star formation activity. These objects are distinct from the population of X-ray dim, 'normal', star forming galaxies now being revealed in deep high-$Z$ surveys (e.g., Lehmer et al. 2008). %(e.g., REFS).

\begin{figure}
\begin{center}
\hspace{-6.5mm}
\vbox{
\psfig{figure=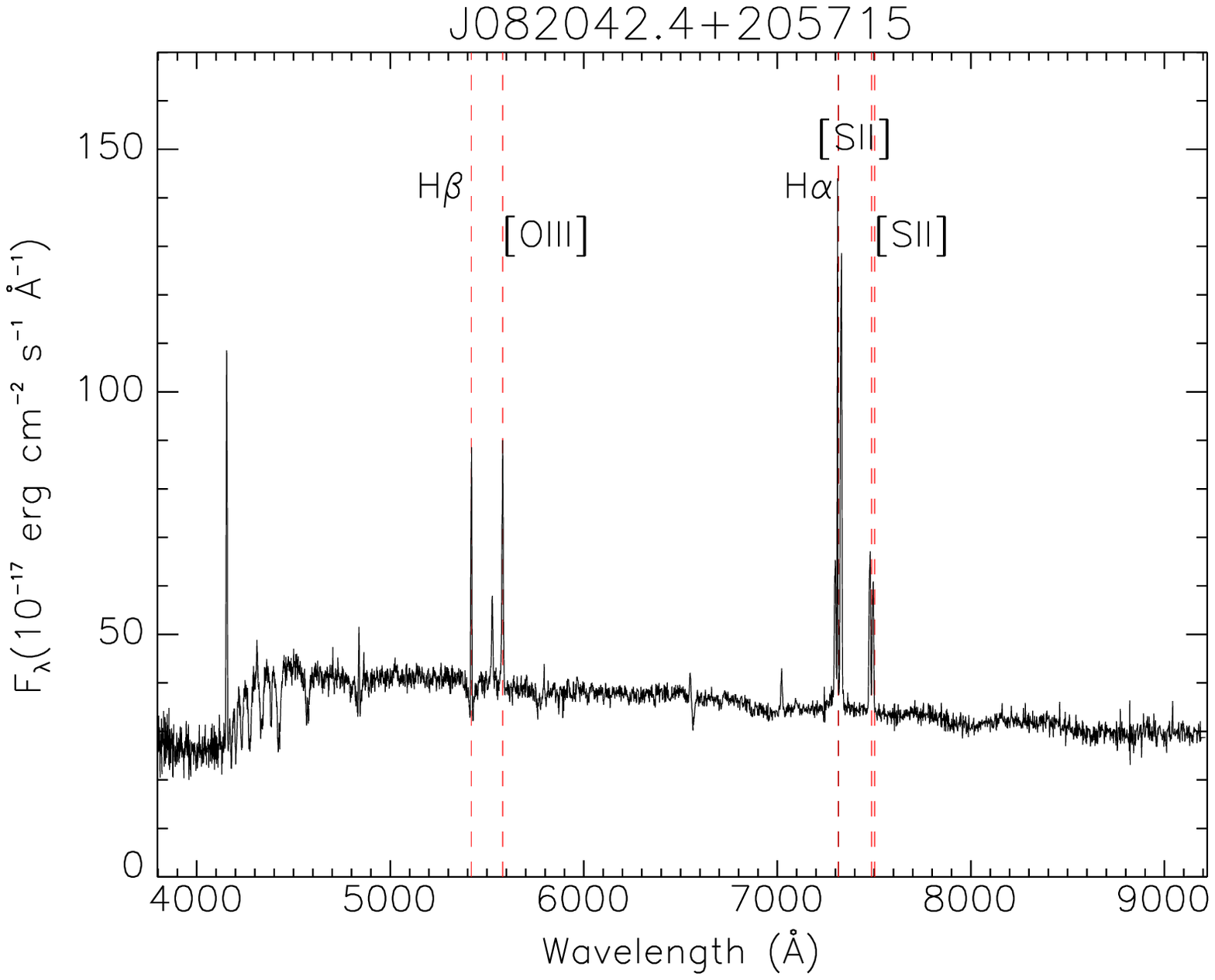,width=90mm}
\vspace{5mm}
\psfig{figure=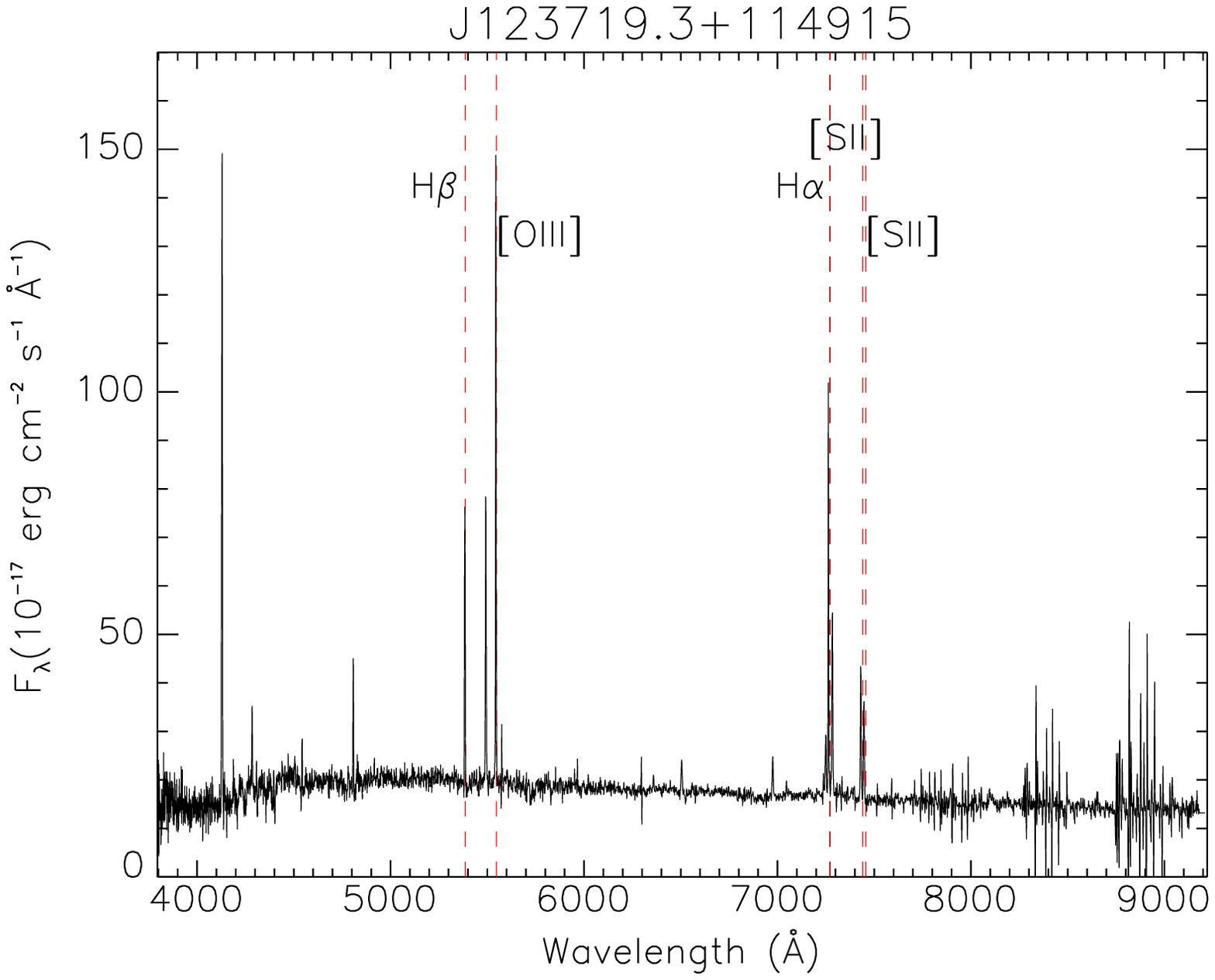,width=90mm}
\vspace{5mm}
\psfig{figure=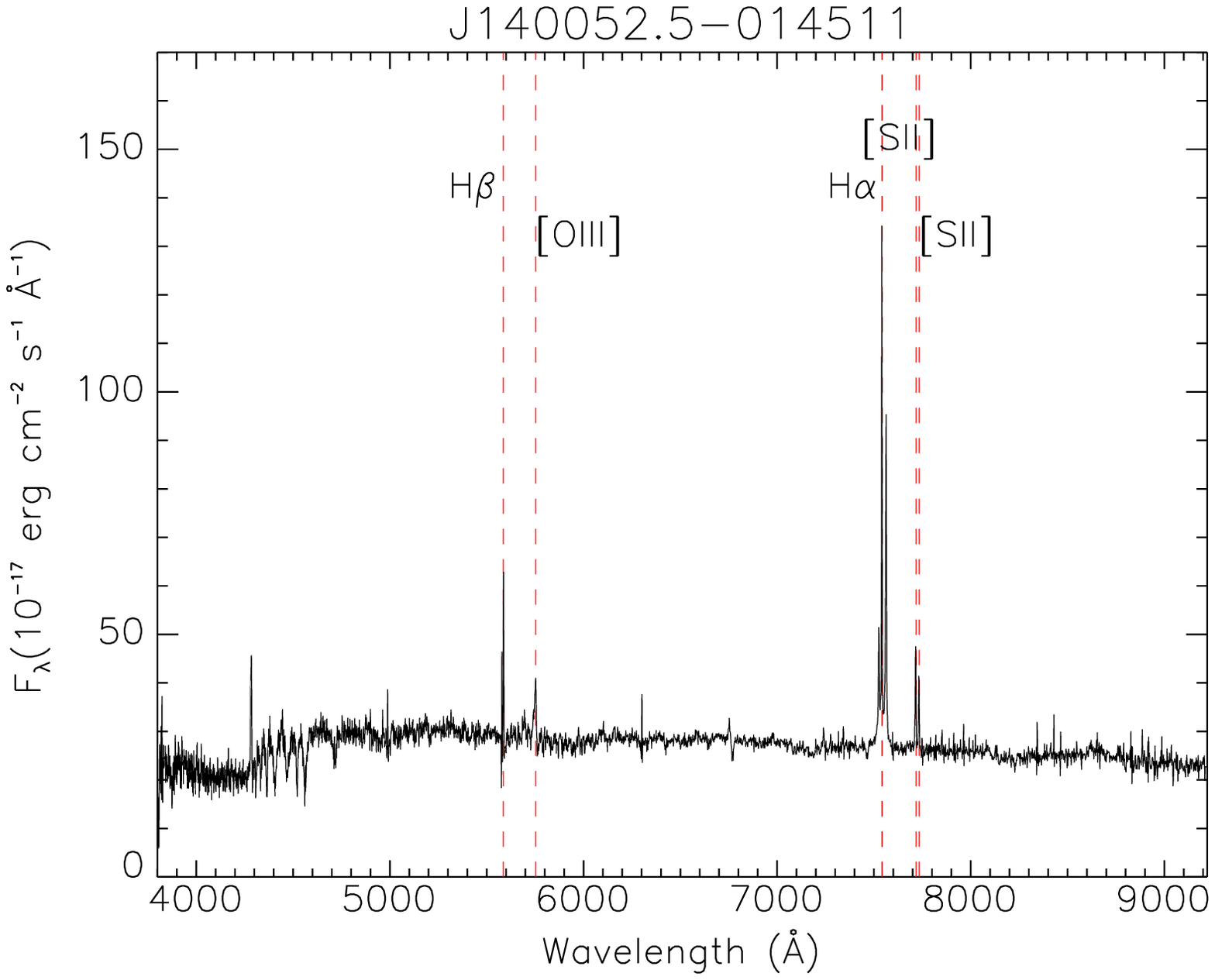,width=90mm}
} 
\caption{SDSS DR7 optical spectra of J082042.4+205715 (top), J123719.3+114915 (middle), and J140052.5-014511 (bottom). None of the spectra show strong signs of AGN activity. Also, signatures for star formation, such as the presence of dominant narrow H{\small $\alpha$} and H{\small $\beta$} emission lines, can be seen within these spectra, indicating the existence of star forming activity within the sources' nuclei.}
\label{fig:optical}
\end{center}
\end{figure}
 
If confirmed as star forming galaxies, then these luminous X-ray
sources will be the most powerful starburst galaxies known (with
$L_{\rm X}\approx10^{42}$--$10^{43}$~erg~s$^{-1}$), exceeding the
X-ray luminosity of the brightest star forming galaxies in the local
Universe by over an order of magnitude (e.g.,\ Moran et~al. 1999;
Zezas et~al. 2001; Lira et~al. 2002; Ptak et~al. 2003). However,
accurately characterising the X-ray emission from these distant
luminous X-ray sources is challenging due to their faint fluxes and
therefore limited X-ray photon statistics. Greater insight into their
nature can be gained by selecting the rare nearby ($z$ $<$ 0.15) examples of these luminous X-ray
sources where the photon statistics are much improved (e.g.,\ Horschemeier et~al. 2005). The advantages of this
approach are twofold; firstly the accurate characterisation of the X-ray emission
via X-ray spectral and variability analyses, and secondly the accurate
characterisation of the optical spectroscopic signatures.

In this paper we present {\it XMM-Newton} data and detailed analyses
of three X-ray luminous galaxies at $z\approx$~0.1. These targets were
selected from the $\approx$~80~deg$^2$ overlap between the pre-release version of the source catalogue of the \xmmn~Serendipitous survey (2XMMp-DR0)\footnote{The 2XMMp catalogue (http://xmmssc-www.star.le.ac.uk/Catalogue/2XMMp/) is a subset of the full 2XMM catalogue (Watson et al., 2009). Most of the details of its production are very similar to the full 2XMM catalogue, as described by Watson et al. (2009).} and the Sloan
Digital Sky Survey data 5 (SDSS-DR5)\footnote{\tt http://www.sdss.org/dr5}. %Three sources were selected based on their lack of optical AGN signatures and their high X-ray luminosities of $L_{\rm X}$ $>$ 10$^{42}$ erg s$^{-1}$. 
The aim of this study is simply to determine whether the X-ray emission is powered by star formation, or by AGN activity. This paper is structured as follows: in section 2, we give  details about the source selection criteria. The X-ray observations used in this study are described in section 3, along with the data reduction details. In section 4, we describe the results of the spatial, spectral, and timing analyses performed on the X-ray data for each source. In section 5, we discuss the results %possibilities of why the sources are optically elusive.
, and finally, we conclude the paper in section 6. Throughout this work, we assume the values $H_{\rm 0} = 71$ km
s$^{-1}$ Mpc$^{-1}$, $\Omega_{\rm M} = 0.27$ and $\Omega_{\rm Vac} =
0.73$.

\section{Source Selection}

As an initial step, we cross-correlated the SDSS DR5 and 2XMMp-DR0 catalogues.
From the large number of cross-matches we then selected X-ray sources
that were identified with galaxies possessing star formation dominated
SDSS optical spectra, with no clear AGN signature. The
criteria used included, for example, strong narrow emission lines, low
[S{\small II}]/H$\alpha$%, [N{\small II}]/H$\alpha$, 
and [O{\small III}]/H$\beta$ ratios, and
underlying continua dominated by young stars. 

\begin{table*}
%\leavemode
\begin{center}
\caption{Catalogued properties of the X-ray luminous starburst galaxies}
\label{tab:tab0}
\begin{tabular}{ccccccccc}
\hline
\multicolumn{2}{c}{Source ID} & Mag $^{a}$ & $z$ & $F_{25}~^{b}$ & $R_{25}~^{c}$ & $A_{\rm B}~^{d}$ &$N_{\rm H}~^{e}$ & Notes $^{f}$ \\
2XMMp... & SDSS... & & & (arcsec) & & (mag) & (10$^{20}$ cm$^{-2}$) & \\ 
\hline
J082042.4+205715 & J082042.46+205714.9 & 16.6g & 0.114 & 9.6 & 1.28 & 0.213 & 4.25 & (1)  \\
J123719.3+114915 & J123719.34+114915.9 & 18.3g & 0.108 & 3.6 & 1.09 & 0.183 & 2.56 & \\
J140052.5-014510 & J140052.57-014510.0 & 17.7g & 0.149 & 5.4 & 1.13 & 0.232 & 3.89 & (2) \\ 
\hline
\end{tabular}
\end{center}
\begin{minipage}{\linewidth}
Notes: {$^a$ Observed SDSS magnitude and filter; $^b$ Semi-major axis of the
$D_{25}$ isophotal ellipse; $^c$ Ratio of $D_{25}$ major and minor
axes; $^d$ Galactic foreground extinction according to Schlegel et
al. (1998); $^e$ Galactic neutral absorbing column from \colden; $^f$
Additional notes: (1) This source was previously detected in the 0.2 --
2 keV X-ray band by {\it ROSAT\/}, at a flux of 7.6$\times10^{-14}$
erg cm$^{-2}$ s$^{-1}$. It was also detected as an infra-red source by
the 2MASS survey with a J-band magnitude of 14.33, and as a radio
source in the FIRST and NVSS surveys at a 1.4 GHz flux density of 4
mJy.  (2) Also detected in the 2dF Galaxy Redshift Survey (Colless et
al. 2001) as 2dFGRS TGN272Z262, confirming the above redshift
measurement. Data for the galaxy properties are taken from the SDSS
catalogue, and the notes result from searches in the NED (http://nedwww.ipac.caltech.edu/) and SIMBAD (http://simbad.u-strasbg.fr/simbad/)
databases.}
\end{minipage}
\end{table*}

\begin{table*}
\begin{center}
\caption{Estimated star formation rates for the target galaxies}
\label{tab:sf}
\begin{tabular}{ccccccccc}
\hline
Source ID & Balmer Decrement & {\it A}$_{\rm V}$$^a$ & {\it L}$_{\rm H\alpha}$$^b$ & SFR$_{\rm H\alpha}$$^c$ & {\it F}$_{\rm 1.4~ GHz}$$^d$ & SFR$_{\rm 1.4~ GHz}$$^e$ & \lx$^f$ & \sfrx$^g$  \\
(2XMMp...)& & (mag) & ($\times$ 10$^{42}$ erg s$^{-1}$) & (M$_{\sun}~$ yr$^{-1}$) & (mJy) & (M$_{\sun}~$ yr$^{-1}$) & ($\times$ 10$^{42}$ erg s$^{-1}$)  & (M$_{\sun}~$ yr$^{-1}$)  \\
\\
\hline
J082042.4+205715 & 6.55 & 2.40 & 3.55 & 28.1 & 4.5$^{i}$ & 110 & 5.4 & 400 \\
J123719.3+114915 & 4.20 & 1.11 & 0.70 & 5.6 & $<$ 4.4$^{ii}$ & $<$ 95 & 2.8 & 200 \\
J140052.5-014510 & 7.68 & 2.86 & 4.21 & 33.4 & $<$ 1.1$^{i}$ & $<$ 50 & 13.9 & 1000 \\
\hline
\end{tabular}
\begin{minipage}{\linewidth}
Note: $^a$ V-band extinction, calculated with the assumption that the non-reddened Balmer ratio should be 2.86; $^b$ De-reddened H{\small $\alpha$} luminosity; $^c$ H{\small $\alpha$} star formation rate as per Doherty et al (2006); $^d$ 1.4 GHz radio flux; $^e$ 1.4 GHz radio star formation rate as per Bressan et al (2002). Note that the radio fluxes recorded in this table for both J123719.3+114915 and J140052.5-014511 are conservative 3$\sigma$ upper limits; $^f$ Initial 0.3 -- 10.0 keV X-ray luminosity measurement based on the 2XMMp catalogue; $^g$ X-ray star formation rates calculated as per Grimm et al. (2003). [$^i$ as observed by the \nvss; $^{ii}$ as observed by the \first survey.]
\end{minipage}
\end{center}
\end{table*}

\begin{table*}
\leavevmode
\begin{center}
\caption{Summary of the observational data used in this paper.}
\label{tab:obs}
\begin{tabular}{cccccccc}
\hline
Source ID & Obs ID & Observation & Instrument $^b$ & Off-axis angle $^c$ & Exposure $^d$ & Detected source  & Light curve  \\
            &        &     date $^a$&                &          &         & counts           & bin size $^{e}$\\
(2XMMp...)            &        &     &                & (arcmins)          &   (ks)       &           & (s)\\
\hline
J082042.4+205715 & 0108860501$^1$ & 15-10-2001 & EPIC & 7.1 & 16.3 & 625$\pm$61 & 2500 \\
J082042.4+205715 & 7925$^1$ & 18-09-2007 & ACIS-I & 9.7 & 48.8 & 278$\pm$19 & 5000 \\
J082042.4+205715 & 0505930301$^{*2}$ & 04-04-2008 & EPIC & 0 & 36.6 & 4405$\pm$133 & 1500 \\
J123719.3+114915 & 0112840101$^{1}$ & 12-06-2003 & EPIC & 4.7 & 14.7 & 185$\pm$55 & 2500 \\
J123719.3+114915 & 9558$^{2}$ & 29-02-2008 & ACIS-S & 0 & 49.3 & 134$\pm$12 & 12500\\
J140052.5-014510 & 0200430901$^{1}$ & 02-07-2004 & EPIC & 11.0 & 7.5 & 571$\pm$41 & 500 \\
J140052.5-014510 & 0505930101$^{*2}$ & 18-10-2007 & EPIC & 0 & 21.4 & 3743$\pm$99 & 1000\\
J140052.5-014510 & 0505930401$^{*2}$ & 10-01-2008 & EPIC & 0 & 16.4 & 1760$\pm$91 & 500\\
J140052.5-014510 & 9557$^{2}$ & 11-03-2008 & ACIS-S & 0 & 49.4 & 1780$\pm$42 & 2500\\
\hline
\end{tabular} \\
\end{center}
\begin{minipage}{\linewidth}
Notes: $^a$ Date in dd-mm-yyyy. $^b$ EPIC=\xmmn EPIC-MOS and pn cameras; ACIS-S =
\chan ACIS-S camera; ACIS-I = \chan ACIS-I camera. $^c$ Angular distance
between the pointing axis for the observation (for {\it XMM-Newton\/}
this is taken from the pn detector axis) and the source. $^d$ The
flaring-compensated exposure time. $^e$ The timing analysis light
curve bin sizes.  $^*$ The targets of these observations are recorded
in the archive as 2XMMp J082042.5+205715, and 2XMMp J140052.6-014511,
as a result of rounding errors by the proposers. $^{1}$ Observations from which the data are archival. $^{2}$ Observations from which the data was obtained especially for this study.
\end{minipage}
\end{table*}

These serendipitous X-ray bright starburst galaxy candidates were then
further filtered to provide a small sample of objects for follow-up
observations in order to conduct a more detailed investigation. The main criteria were: (i) unusually high apparent
X-ray luminosity for a starburst galaxy, $L_{\rm X} > 10^{42} \rm
~erg~s^{-1}$, and (ii) an estimated on-axis EPIC-pn count rate in
excess of 0.01 ct s$^{-1}$ for better photon statistics. Nineteen sources were found to match the first criterion. Of these sources, three were identified
matching the second criterion: 2XMMp J082042.4+205715, 2XMMp
J123719.3+114915, and 2XMMp J140052.5-014510. The characteristics
of these three objects are presented in Table~\ref{tab:tab0}. Figure~\ref{fig:bpt} shows the emission-line diagnostic diagrams representing the [S{\small II}]/H$\alpha$ versus [O{\small III}]/H$\beta$ and [N{\small II}]/H$\alpha$ versus [O{\small III}]/H$\beta$ ratios of a large sample of galaxies, selected from the SDSS DR7 catalogue. Included in this sample are the ratios of our three galaxies, represented by the three filled squares in both panels. In the case of the [S{\small II}]/H$\alpha$ diagnostic, all three of the sources' flux ratios lie well within the starburst region of the diagram. In the [N{\small II}]/H$\alpha$ diagram, the three sources' ratios lie within the diagram's Transition region. %None of the sources' flux ratios lie within the AGN region of the diagram (although two sources' ratios lie very close). They are all positioned in the diagram's ambiguous LINER/Transition region.
%In 
%We also looked at the case of the [N{\small II}]/H$\alpha$ versus [O{\small} III]/H$\beta$ diagnostic, and found that the three sources' ratios lie within the diagram's Transition region. 
This, however, does not indicate a definite presence of an AGN. For example, a number of pure star forming galaxies such as NGC 7714, Mrk 711, and UM 304 have flux ratios that pertain to those of Transition sources (Terlevitch et al. 1991). No evidence has been reported on the presence of AGNs within these galaxies.   
The filled circles represent the remaining 16 sources that were selected from the SDSS catalogue based on high X-ray luminosity.

We obtained SDSS optical data for the objects from the SDSS DR7 catalogue\footnote{The observations were taken in 2002 and are associated with the SDSS-DR5 release. The reduction of the spectra was done in 2008, and is associated with SDSS-DR7 (http://www.sdss.org/dr7/)}. The sources' SDSS optical spectra can be seen in Fig.~\ref{fig:optical}. Optical AGN signatures (i.e., emission lines from broad and narrow line regions near to an AGN) are lacking in these spectra. However, signs of star formation, such as the presence of strong H{\small $\alpha$} and H{\small $\beta$} lines, are seen. Along with the optical data, radio data were obtained from the NRAO \vla Sky Survey (\nvss)\footnote{\tt http://www.cv.nrao.edu/nvss/} and the \vla Faint Images of the Radio Sky at Twenty-Centimeters (\first) survey\footnote{\tt http://sundog.stsci.edu/} for each source. Using both the optical and radio data, we estimated the sources' H$\alpha$ and 1.4 GHz radio fluxes, and converted them to star formation rates as per Doherty et al. (2006) and Bressan et al (2002), respectively. We also estimate the sources' X-ray star formation rates as per Grimm et al. (2003). Estimated star formation rates can be seen in Table~\ref{tab:sf}. In all three sources, a moderate H$\alpha$ star formation rate is seen, ranging from $\sim$ 6 to $\sim$ 30 M$_{\sun}~$ yr$^{-1}$. Two of the three radio star formation values are calculated using conservative 3$\sigma$ upper limits to radio flux, and therefore only represent upper limits to star formation rates, with SFR$_{\rm 1.4GHz}$ $<$ 95 and $<$ 50 M$_{\sun}~$ yr$^{-1}$ for J123719.3+114915 and J140052.5-014511, respectively. The one remaining star formation rate is relatively high, at 110 M$_{\sun}~$ yr$^{-1}$ for J082042.4+205715. The X-ray estimates are a lot higher than these values, at 400, 200, and 1000 M$_{\sun}~$ yr$^{-1}$. These extremely high values were calculated on the assumption that the X-ray emission from these galaxies is powered by star formation alone. This suggests that non-stellar activity is dominating the X-ray emission. We investigate this possibility in later sections of the paper. 

Because the sources are off-axis in the serendipitous observations, these data are not good enough to yield high quality photon statistics. Therefore, new \xmmn observations were sought and obtained for two of the objects
(PI: Roberts); two of the three were also awarded new observations on
\chan (PI: Gelbord) with the third already being observed at the edge
of the NGC 2563 group field-of-view. Details of the full set of \xmmn
and \chan observations of the three objects are presented in
Table~\ref{tab:obs}. 

\section{Observations and Data Reduction}

A total of nine X-ray observations are examined in this analysis. The archival data sets were either downloaded from the \xmmn
Science Archive\footnote{\tt http://xmm.esac.esa.int/xsa/} or the
\chan Data Archive\footnote{\tt http://cxc.harvard.edu/cda/}. The
three targets were initially identified at a variety of off-axis angles in three
separate serendipitous \xmmn detections (see Table 3).
J082042.4+205715 and J140052.5-014510 were subsequently observed
on-axis as the target of follow-up \xmmn observations, with the latter
observed on two separate occasions.  All observations were taken in a
full frame imaging mode, although in the \xmmn observation 0505930101, only the pn and MOS1 detectors obtained sufficient data for the analysis. This is because the pn had a full active buffer during the observation, which resulted in the data begin lost from the MOS2 detector. The overall exposure of MOS2 was $\sim$ 5 ks as a result, thus yielding a net photon count of $\sim$ 20. % This is because the MOS2 exposure time is $\sim$ 5 ks, yielding $<$ 20 source photons in this exposure.    

Reduction of the \xmmn data was performed using \xmmn \sas version
8.0.0 software. Contamination by background flaring was present in all
observations. In order to remove this, good time interval (GTI) files
were created by excluding periods during which the count rate in the
10 - 15 keV full field pn light curves exceeded 1 ct s$^{-1}$, and the
data were filtered to remove these periods in the subsequent analyses.

Spectral data were extracted from circular regions of radius 45
arcseconds, centred on each of the source positions, with only two
exceptions.  In observation 0112840101, J123719.3+114915 is located
immediately adjacent to a chip gap in the MOS1 data, forcing us to
obtain the data in an aperture of only 12.5 arcseconds radius. The resulting aperture encircled only 50\% the energy of the source. This loss of area was compensated for by the corresponding source ancilliary matrix file (see below). In the
case of the 0505930301 data, J082042.4+205715 was sandwiched between a
nearby source -- not present in the 0108860501 data -- and a pn chip
gap, causing the aperture to be shrunk to 30 arcseconds radius (a
slight offset of the aperture allowed a full 45 arcsecond radius
region to be used in the MOS data).  Corresponding background files
were extracted from source-free circular regions with radii at least
1.5 times greater than source extraction regions, or multiple regions
with a summed area equivalent to this.  They were positioned on the
same chip as, and in close proximity to, the source regions. The
spectral data files were extracted with FLAG = 0 and either PATTERN
$\le$ 4 for pn or PATTERN $\le$12 for MOS, using standard \xmmn
tasks. Response matrix and ancillary response files were also created
as part of these tasks.  In all but one of the resulting spectral
files the data were grouped into bin sizes of at least 20 counts per
bin. The spectra from 0112840101 were left unbinned as the overall
number of photon counts (185 $\pm$ 55 cts) was too low to be compatible with model
fitting via standard $\chi^2$ analysis. Instead, Cash statistical
analyses (Cash 1979) were utilised on the spectral data (see Section 4.3).

Source and background light curves were extracted in the standard
manner to facilitate timing analysis of the \xmmn data. This used the
same regions from which the spectral files were extracted. For the
analysis of this data, we set a minimum signal to noise ratio (S/N)
requirement of $\geq$ 5 per bin. In order to accommodate this, the
light curve bin sizes were set according to the source photon count
rate. Table~\ref{tab:obs} contains details of the temporal binning
corresponding to individual observations. In order to correct for GTI
gaps (due to background flaring) that left substantial time periods
missing within individual light curve bins - and hence gave erroneous
count rates - we created new GTI files on the same binning as the
light curves.  These were used when extracting the light curves, and
effectively removed the time bins associated with flaring, leaving only bins
containing reliable count rates.

\begin{figure*}
\begin{center}
\hspace{-0.1cm}
\hbox{\fbox{\psfig{figure=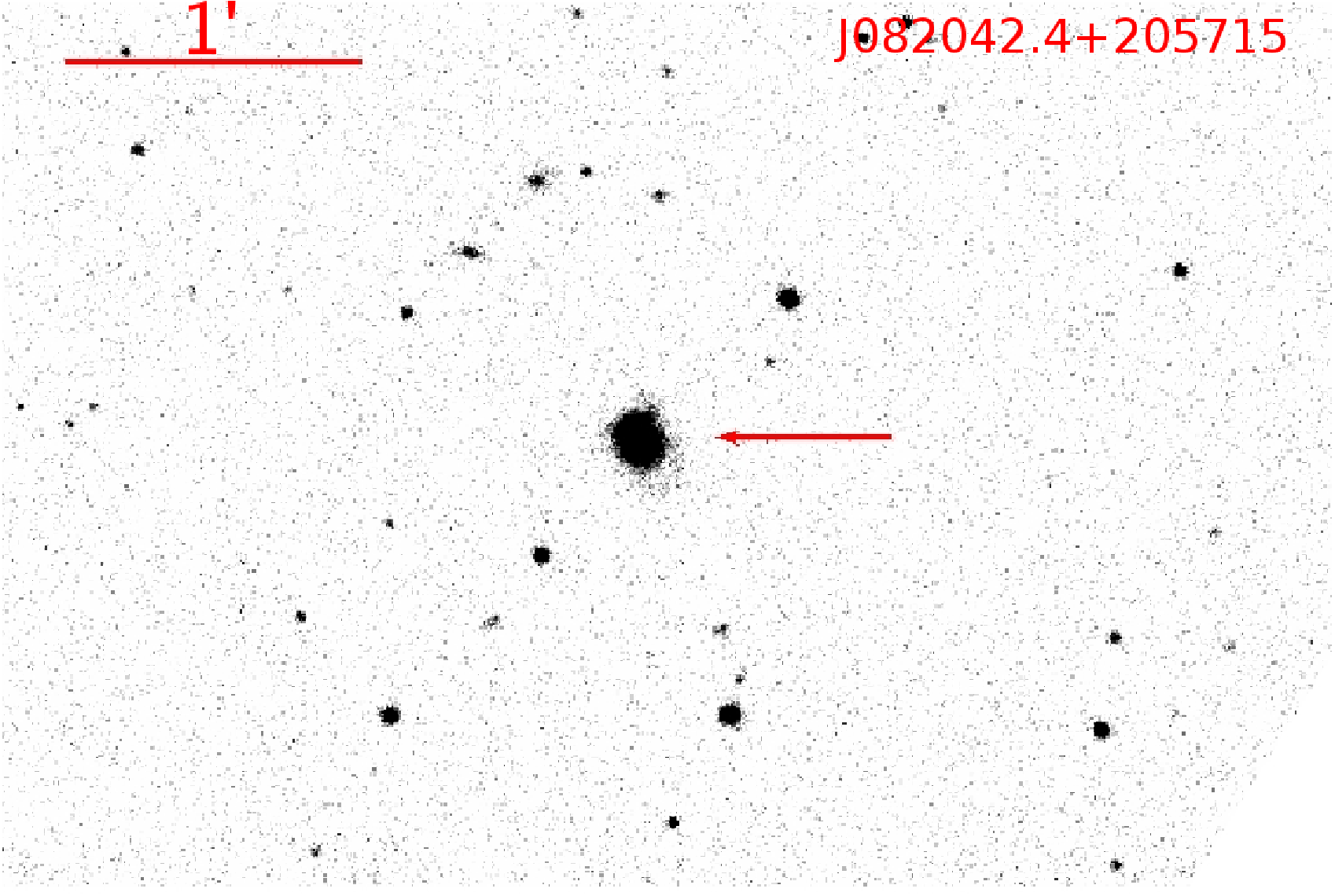,width=5cm,height=4cm} }
%\hspace{0.1cm}
\fbox{\psfig{figure=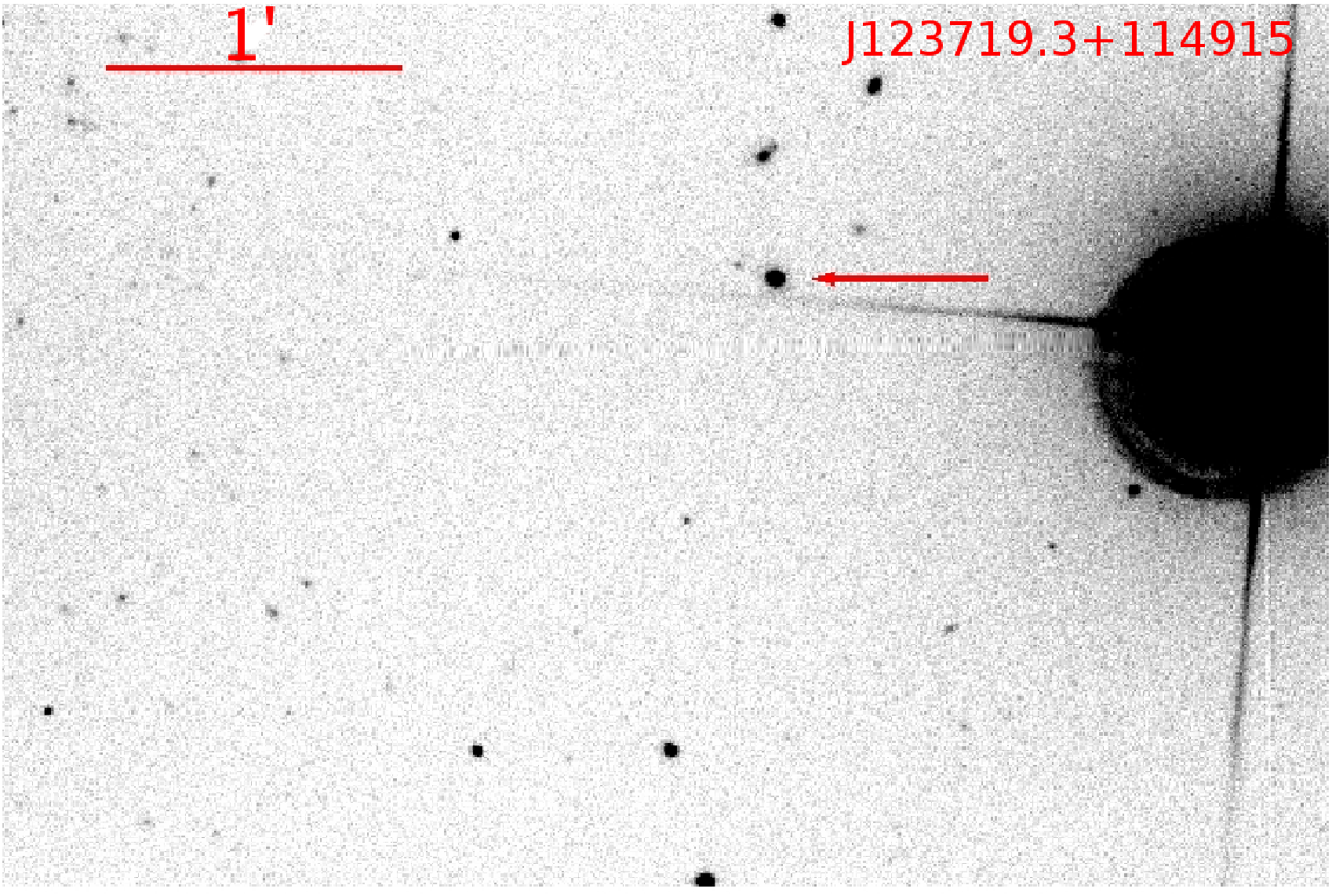,width=5cm,height=4cm}}
%\hspace{0.1cm}
\fbox{\psfig{figure=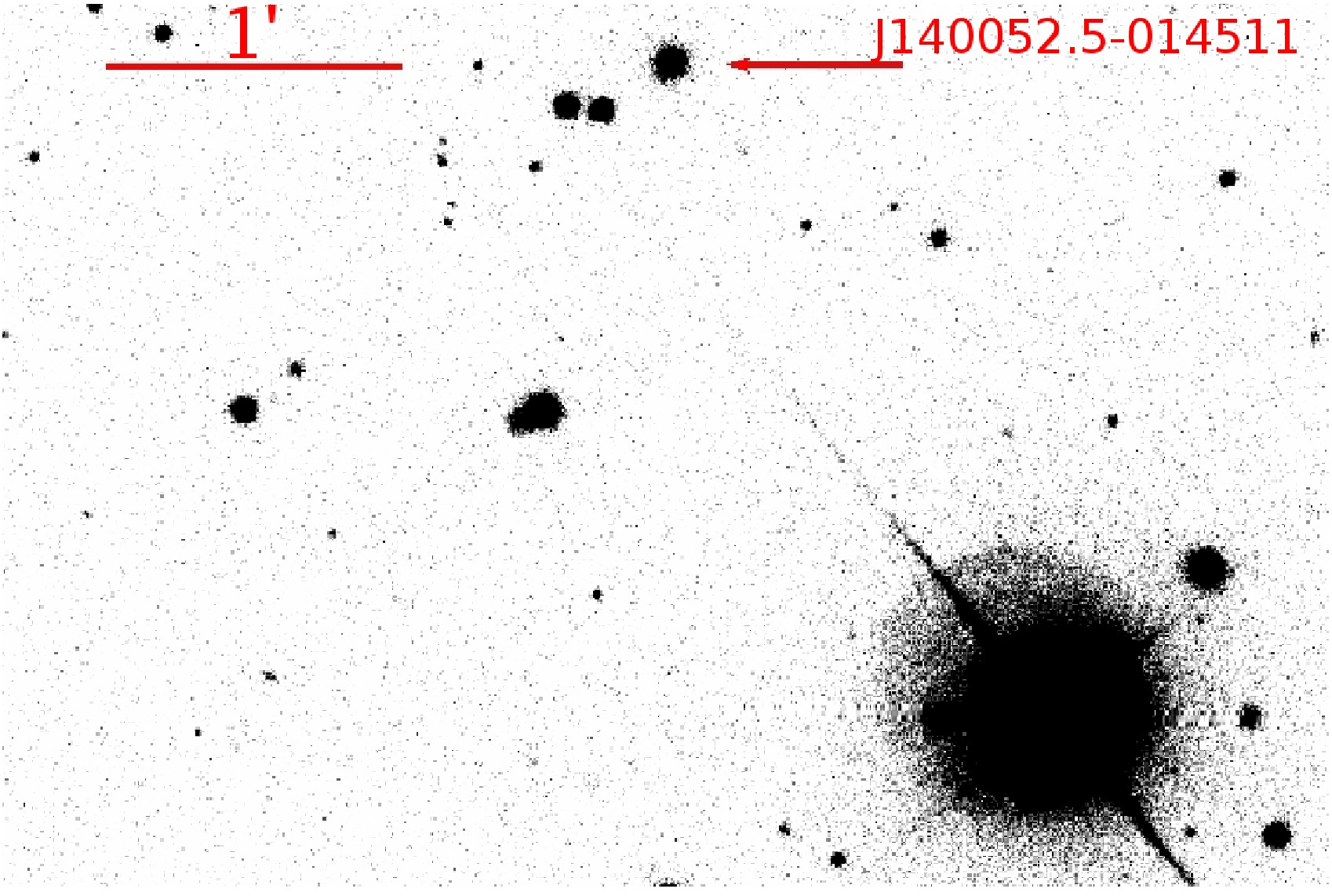,width=5cm,height=4cm}}}
\vspace{0.0cm}
%\hspace{-0.05cm}
\hbox{\fbox{\psfig{figure=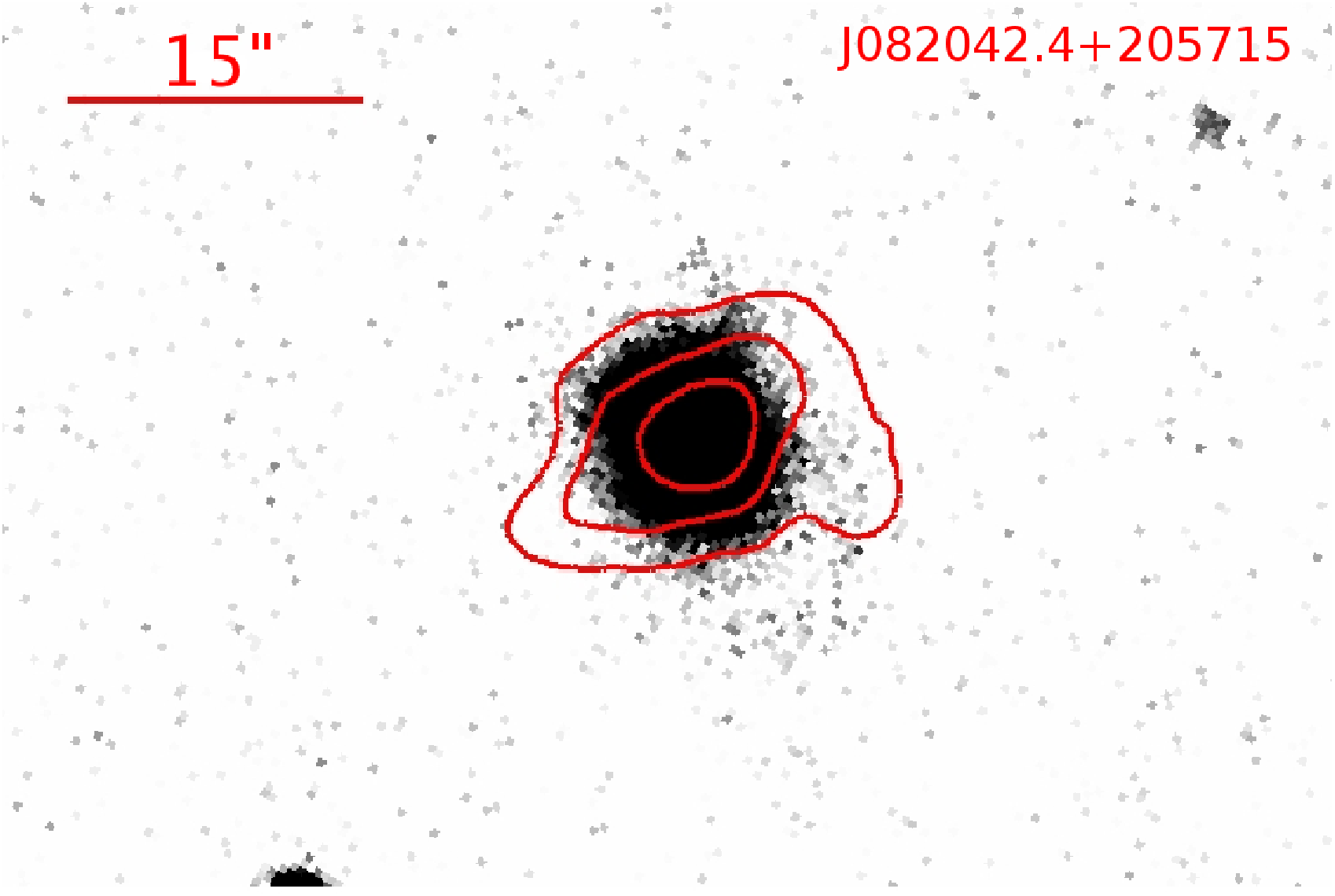,width=5cm,height=4cm}}
%\hspace{-0.1cm}
\fbox{\psfig{figure=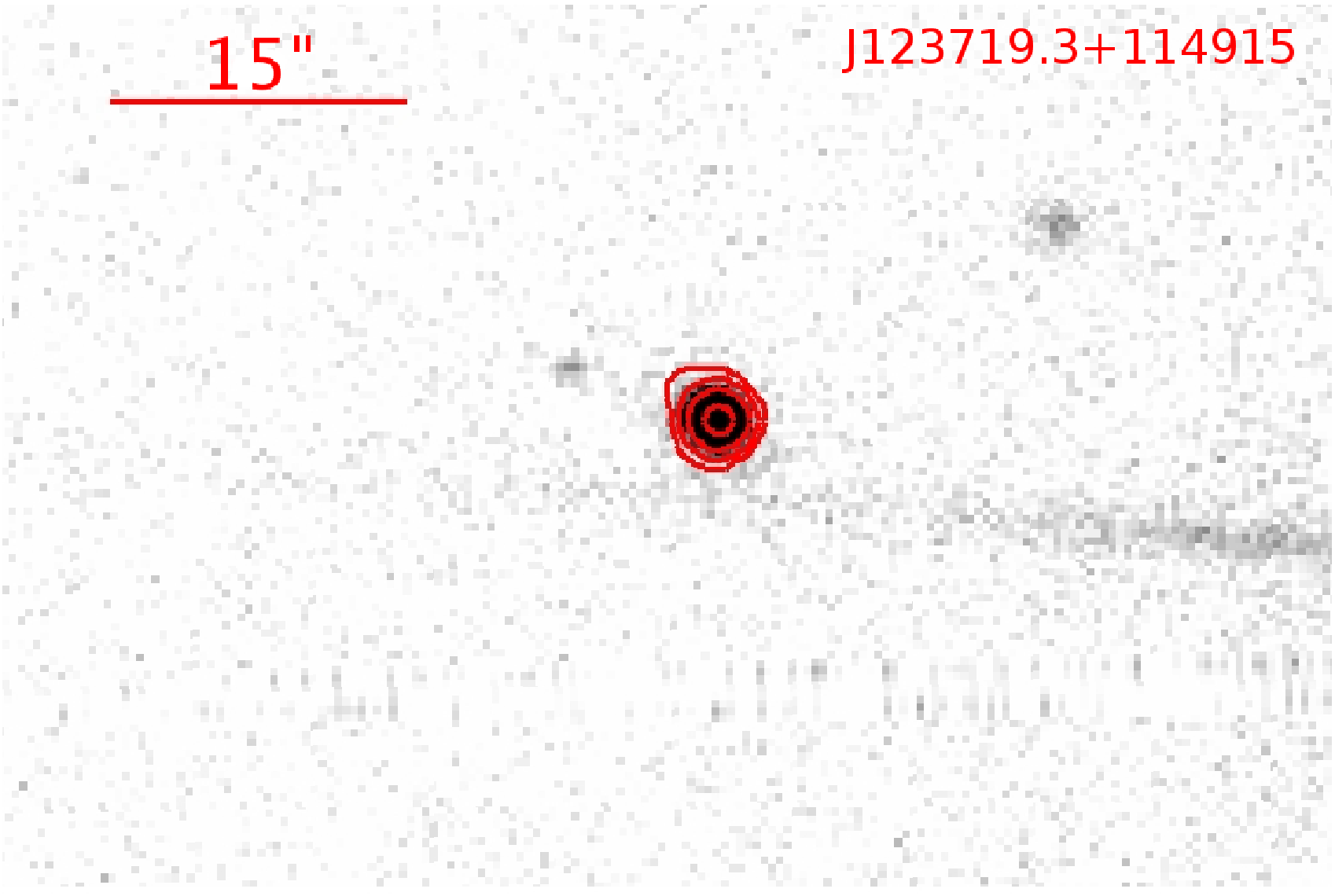,width=5cm,height=4cm}}
%\hspace{-0.1cm}
\fbox{\psfig{figure=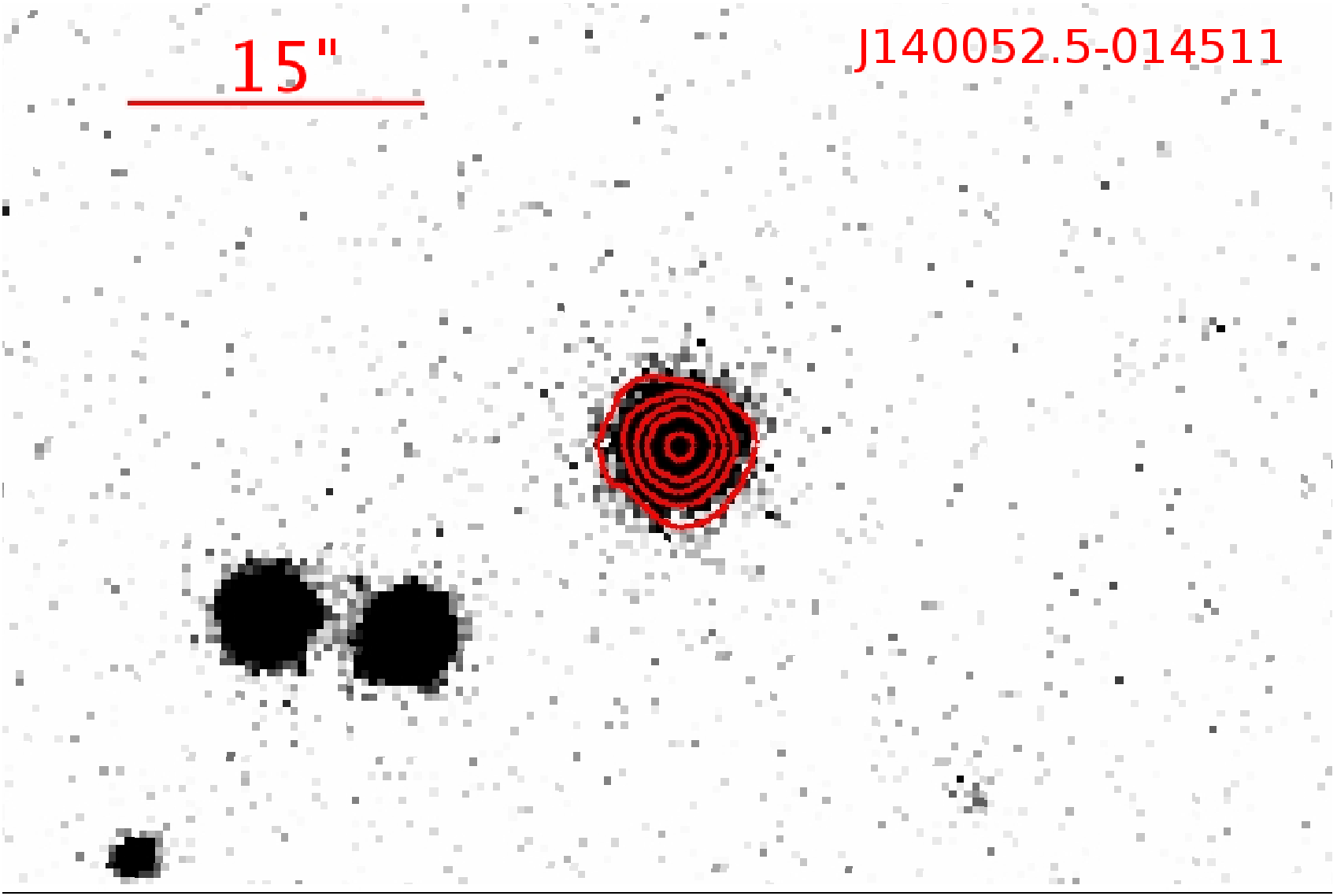,width=5cm,height=4cm}}}
\caption{{\it Top row\/}: From left to right, SDSS R-band 
images containing the three sources J082042.4+205715,
J123719.3+114915, and J140052.5-014510 respectively.  The objects that
are the focus of this paper are indicated. J082042.4+205715 is located
$\sim 7$ arcminutes to the south of NGC 2563, the central galaxy of
the NGC 2563 group of galaxies (e.g. Mulchaey et al. 2003), although
our object is clearly located in the background to this nearby ($d
\sim 60$ Mpc) group.  The other objects are in emptier fields,
although both are located close to the line-of-sight to nearby, bright
K0 stars.  J123719.3+114915 is $\sim 1.5$ arcminutes east of the
$m_{\rm V} \sim 8$ star HD 109771, and J140052.5-014510 $\sim 2.5$
arcminutes north-east of the $m_{\rm V} \sim 9.4$ star HD 122238. The
double source to the immediate south-east of J140052.5-014510 is a pair of point-like objects with equal magnitudes ($g =
17.2$), recorded as J140053.49-014520.6 and J140053.97-014519.6 in the
SDSS catalogue. According to the SDSS DR7 catalogue, these two objects have photometric redshifts of $\sim$ 3$\times$ 10$^{-4}$ and 1 $\times$ 10$^{-3}$, respectively. {\it Bottom row\/}: Close up images of
the three sources, in the same order as above, overlaid with X-ray
contour plots. The contour levels start at 0.09 count
pixel$^{-1}$, with each
successive contour representing a count density a factor $\sim 2.6, 4$
and 5.2 higher than the last (left -- right across the three images).  The optical data are
displayed as above, and the X-ray data has been smoothed with a
1-pixel Gaussian kernel for display purposes.}
\label{fig:figrad}
\end{center}
\end{figure*}

Unlike the \xmmn observations, only one of the three sources was
detected serendipitously with {\it Chandra\/} (again, see Table~\ref{tab:obs}). This source,
J082042.4+205715, was detected off-axis by the ACIS-I instrument with
the camera in the full frame, timed exposure mode. The remaining two
sources were observed by the ACIS-S instrument in two separate on-axis
follow-up observations, again with the camera in a full frame, timed
exposure mode.  All of the data were obtained in the very faint
telemetry format.

The \chan data was reduced using \ciao software (version 4.1.1),
adopting standard methods.  Spectra were extracted using the
\ciao \textit{specextract} routine.  In the subsequent spectral
analysis the spectral data files for observations 7925 and 9558 were
left unbinned due to the low quality of the data and, as with the
lower quality \xmmn data, Cash statistical analysis was performed. The
one remaining observation, 9557, possessed sufficient source counts to
permit binning to at least 20 counts per bin and model fitting using
$\chi^2$ statistics. As per the timing analysis with \xmmn data, \chan
light curves were extracted from apertures similar to those used to
extract the spectra. The bin sizes for the light curves were again set
using the S/N $\geq$ 5 requirement (also see Table~\ref{tab:obs} for
the \chan light curve binning).

\section{Results}

\subsection{Point Source Analysis}
	
A first, key test of whether the X-ray emission of the selected
objects originates in an AGN or a starburst region, is whether the
X-ray emission appears point-like at the highest available spatial
resolution.  If extension is present, this would confirm that there is
a spatially extended X-ray emission region, consistent with at least
part of the X-ray emission originating in a large starburst region.
Such a test is possible for our objects as high spatial resolution ($<$ 1 arcsec/pixel)
\chan data are available for all three of them.  We show both the
identification of the counterpart starburst galaxy on an SDSS R-band field, and a close-up of the individual galaxy with \chan
X-ray emission contours overlaid, in Fig.~\ref{fig:figrad}.  Note that
the obvious extension of the contours to J082042.4+205715 can, to
first order, be explained by point spread function (PSF) degradation
due to its off-axis location in the \chan ACIS-I field-of-view
(cf. Table~\ref{tab:obs}).  The other objects appear compact, consistent with a
point-like source on axis.

In order to best determine whether the sources are truly point-like or
extended, it is necessary to undertake a detailed analysis of the PSF
of each of the sources. In order to do this, we first simulated the PSF
of point-like sources with the same flux and off-axis angles as our
objects.  These simulations require the use of both the \chan ray
tracer, \chart\footnote{\tt http://cxc.harvard.edu/chart}, and the
{\sc MARX}\footnote{\tt http://space.mit.edu/CXC/MARX/} simulator.
Firstly, we use \chart to produce a series of simulated monochromatic
PSFs, ranging over the 0.3 -- 10 keV range, with appropriate fluxes and
positions with respect to the detector axis.  These PSFs are the
result of ray-tracing through the \chan HRMA X-ray optics, and still
require projecting onto the detector plane (taking account of the
detector response).  This functionality is provided by {\sc marx},
which we therefore use to complete the simulation process.  We then
compare these simulated PSFs to the actual data using the \ciao {\it
srcextent\/} script\footnote{\tt
http://cxc.harvard.edu/ciao/threads/srcextent/}. This uses a Mexican
Hat Optimisation algorithm to determine a 'size' -- effectively an
average radius -- for each PSF, and directly compares the two. We
show the results from {\it srcextent\/} in Table~\ref{tab:tabpsf} for
the real data, and the simulated monochromatic PSF at 1 keV.

\begin{table}
\begin{center}
\caption{Observed source PSF and simulated (1 keV) PSF details from
{\it srcextent\/}.}
\label{tab:tabpsf}
\begin{tabular}{ccccc}
\hline
Source  & Obs. & PSF & Source & Source \\
ID & ID & size$^{a}$ & size$^{a}$ & size \\
(2XMMp...)& & (arcsec) & (arcsec) & (kpc) \\
\hline
J082042.4+205715 & 7925 & 2.91$\pm$0.35 & 2.55$^{+0.29}_{-0.30}$ & 6.49$^{+0.74}_{-0.76}$ \\
J123719.3+114915 & 9558 & 0.49$\pm$0.04 & 0.47$^{+0.06}_{-0.07}$ & 1.20$^{+0.15}_{-0.18}$ \\
J140052.5-014510 & 9557 & 0.50$^{+0.01}_{-0.02}$ & 0.45$\pm$0.03 & 1.15$\pm$0.08 \\
\hline
\end{tabular}
\begin{minipage}{\linewidth}
Note: $^a$ The Mexican hat optimisation algorithm employed by {\it
srcextent\/} fits a 2-D elliptical Gaussian to the PSF; the 'size'
quoted is calculated as $\sqrt{(1/2) (a^2+b^2)}$, where $a$ and $b$ are
the semi-major and semi-minor axes of the elliptical Gaussian
respectively.  The errors are quoted at the 90\% confidence level.
\end{minipage}
\end{center}
\end{table}

\begin{figure}
\begin{center}
\includegraphics[width=8.4cm,angle=0]{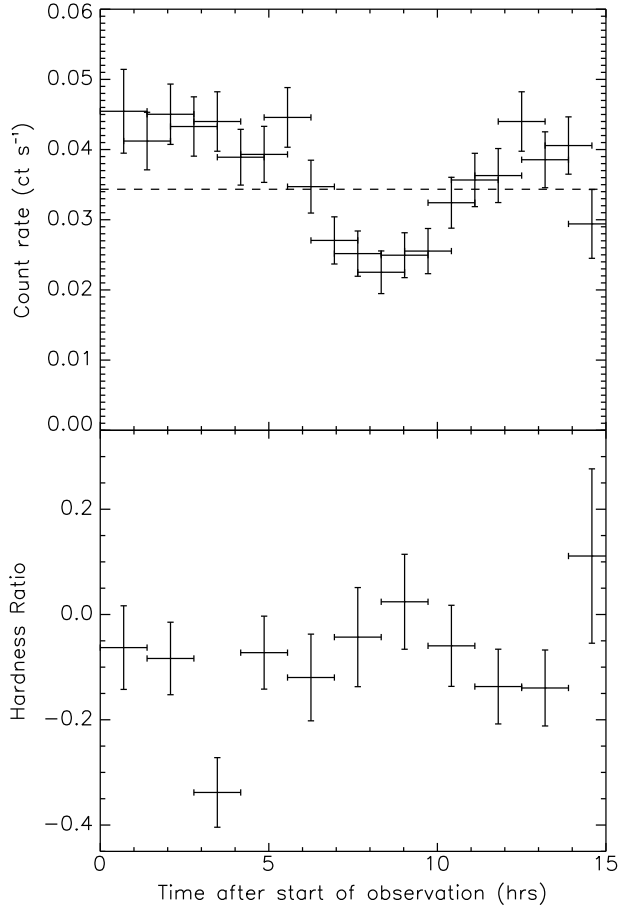}
%\includegraphics[width=8.4cm,angle=0]{acis_lc_new.ps}
%\caption{Light curve of the ACIS-S observation of J140052.5-014510
%(Obs ID 9557).  The horizontal dashed line represents the average
%count rate over the whole observation.  A 'dip' in the count rate is
%clearly visible between $\sim 20$ and 40 ks into the observation.}
%\label{fig:fig2}
%\end{center}
%\end{figure}

%\begin{figure}
%\begin{center}
%\includegraphics[width=8.4cm,height=17cm,angle=0]{9557_HR_new.ps}
\caption{{\it Upper panel:} Light curve of the ACIS-S observation of J140052.5-014510
(Obs ID 9557). The horizontal dashed line represents the best fitting constant
count rate over the whole observation.  A 'dip' in the count rate is
clearly visible between $\sim 20$ and 40 ks into the observation. {\it Lower panel:} Hardness ratio as a function of time after the start of the
observation, for the same observation.}
%\label{fig:fig3}
\label{fig:fig2}
\end{center}
\end{figure}

As Table~\ref{tab:tabpsf} shows, the observational data does not
appear extended with respect to the simulated data at 1 keV (chosen as
an illustrative example as it is close to the peak of the \chan
effective area). Indeed, none of the simulated PSFs (over the typical
ACIS energy range) appeared significantly narrower than the
observational data. [In fact, we note that in most cases the
simulated PSFs appeared slightly broader, although the PSFs generally
agreed at about the 90\% error level.] From the measured extents, we have deduced that the physical sizes of the X-ray sources are constrained to within $\sim$ 6.5 kpc for J082042.4, and to within $\sim$ 1.2 kpc for both J123719.3 and J140052.5. It is worth noting that the compact nature of at least two of these sources does not rule out the presence of star formation within the galaxies' nuclei. It is possible for compact nuclear star formation to be occuring on less than kpc scales, given the distances of the sources. 

\subsection{Intra-observational variability}

A second key distinction between AGN activity and large star formation regions
can be made on the basis of their short-term variability.  A large,
spatially extended region composed of many individual X-ray sources
will not vary coherently on short timescales (i.e. within a single
X-ray observation, typically less than a day). A compact object like
an AGN, on the other hand, may readily vary on timescales of hours
(e.g. Lawrence et al. 1987; Markowitz et al. 2003; Smith \& Vaughan
2007). We therefore analysed the light curve of each observation to
search for short-term variability.

Using the background-subtracted light curves, we employed a $\chi^2$ fit to test whether the X-ray data are variable against a constant count rate hypothesis. To improve statistics, the EPIC-pn and MOS data were
co-added for each observation before testing. All but one of the
datasets were found to show no strong evidence for short-term
variability, consistent with a constant count rate during each
observation. The exception was observation 9557 for J140052.5-014510,
which returned a null hypothesis probability of $<$ 0.001 for a
constant count rate.  We show the light curve for this observation in
the upper panel of Fig.~\ref{fig:fig2}.  In it, the count rate can be seen to drop by a
factor of $\sim$ 2 and then rise back to its previous value within a
matter of $\sim$ 6 hours.  This implies that at least 50\% of the
X-ray emission from this source originates in a region that is no more
than $\sim 6$ light-hours across, consistent with the central regions
of an AGN.

We investigate the nature of this 'dip' in the light curve further in the lower panel of
Fig.~\ref{fig:fig2}, where we display the evolution of the hardness
ratio during the count rate dip.  To calculate this, we split the data
into two bands, a 0.3 -- 1.2 keV soft band ($S$) and a 1.2 -- 10 keV
hard band ($H$), and derived light curves in both, on twice the
binning of the original light curve to maintain good statistics.  We
then combined these into a hardness ratio as $HR = (H-S)/(H+S)$, with
errors calculated as per Ciliegi et al. (1997). The figure
shows that, other than a significant softening in one bin (between 10
and 15 ks into the observation), the hardness of the data remains
roughly constant.  This includes the period of the dip (roughly 20 -
40 ks) -- although the data points suggest a slight hardening in this
period, it is not statistically significant.  This lack of strong
spectral hardening implies that the dip is probably not originating in
a simple increase in absorbing column, lasting for several hours.
Instead, it might have one of two alternative origins; either it is an
intrinsic fluctuation in the output of the compact region, or it is
the result of a Compton-thick medium passing in front of
the X-ray emission region and (at least) partially obscuring it.

\subsection{Spectral Results}

%\begin{table}
%\begin{center}
%\caption{Statistics used in spectral analysis}
%\label{tab:stats}
%\begin{tabular}{ccc}
%\hline
%Source ID & Obs. ID & Statistics\\
%\hline
%J082042.4+205715 & 0108860501 & $\chi^2$ \\
%                 & 7925       & Cash \\
%                 & 0505930301 & $\chi^2$ \\
%J123719.3+114915 & 0112840101 & Cash \\
%                 & 9558       & Cash \\ 
%J140052.5-014511 & 0200430901 & $\chi^2$  \\
%                 & 0505930101 & $\chi^2$ \\
%                 & 0505930401 & $\chi^2$ \\
%                 & 9557       & $\chi^2$ \\ 
%\hline
%\end{tabular}
%\end{center}
%\end{table}

\begin{table*}
\centering
\caption{Results of fitting the spectral data with an absorbed
power-law model.}
\label{tab:PLfits}
\begin{tabular}{ccccccc}
\hline
Source/Obs ID	& $N_{\rm H}$	& $\Gamma$	& $\chi^2$/dof	& C-stat/\%	& EPIC-pn flux	& ACIS flux \\
(1)	& (2)	& (3)	& (4)	& (5)	& (6)	& (7) \\
\hline
{\bf 2XMMp J082042.4+205715} \\
0108860501	& $< 10.9$	& $1.62^{+0.22}_{-0.19}$	& 36.5/42	&	& $1.29^{+0.31}_{-0.13}$	& \\
7925 		& $< 8.2$	& $1.94^{+0.15}_{-0.25}$ 	& 	& 554.1/99	& 	& $1.63^{+0.02}_{-0.14}$ \\
0505930301	& $< 4.5$	& $1.73^{+0.08}_{-0.04}$	& 209/220		& 	& $2.23^{+0.11}_{-0.12}$ \\
{\bf 2XMMp J123719.3+114915}& \\
0112840101 	& $< 10.3$	& $0.48^{+0.29}_{-0.26}$	& 	& 1180/69		& $1.21^{+0.26}_{-0.19}$  \\
9558 	& $82.0^{+82.1}_{-50.2}$	& $1.12^{+0.55}_{-0.48}$	&	& 343.8/50	& 	& $0.44^{+0.07}_{-0.17}$ \\
{\bf 2XMMp J140052.5-014510}& \\
0200430901 & $< 3.5$	& $2.31^{+0.22}_{-0.18}$	& 25.1/24	& &$1.95^{+0.16}_{-0.35}$	 \\
0505930101 & $< 4.1$	& $2.59^{+0.15}_{-0.11}$	& 168/169	& & $2.27^{+0.14}_{-0.08}$	\\
0505930401 & $4.5^{+5.1}_{-3.8}$	& $2.74^{+0.28}_{-0.20}$	& 111/97	& & $1.13^{+0.08}_{-0.11}$	\\
9557		& $<5.2$		& $2.06^{+0.13}_{-0.10}$	& 105/85	& & & $2.34^{+0.16}_{-0.1}$	\\
\hline
\end{tabular}
\begin{minipage}{\linewidth}
Columns: (1) Source name, and observation ID for the spectral data.  (2)
Absorption column density intrinsic to the source ($\times 10^{20}$
cm$^{-2}$).  An additional foreground Galactic column (as per
Table~\ref{tab:tab0}) is also applied and held fixed in the fit.  (3)
Power-law photon index.  (4) $\chi^2$ value and number of degrees of
freedom for the best fitting model.  (5) C-statistic for best fitting
model (in the cases where maximum likelihood fitting was used due to
the low data quality), with the percentage of Monte-Carlo simulations
with a value less than that obtained in the fit to the model.  (6) \&
(7) Observed 0.3 -- 10 keV band flux ($\times 10^{-13}$ erg
cm$^{-2}$s$^{-1}$).
\end{minipage}
\end{table*}

The third diagnostic test we apply to the data is to analyse the X-ray
spectrum of each dataset.  Although AGN are now known to display very
complex spectral phenomenology in the soft X-ray regime including warm
absorbers (e.g. Blustin et al. 2005) and soft excesses
(e.g. Gierli{\'n}ski \& Done 2004), in the absence of high quality
data their spectra are still commonly fitted with simple power-law
models (e.g. in deep surveys, Alexander et al. 2005).  Starburst
regions also have complex spectra (see e.g. Pietsch et al. 2001; Read
\& Stevens 2002), but, in addition to a strong continuum, these are
dominated by relatively narrow emission lines from hydrogen-like and
helium-like charge states of the abundant low-$Z$ elements that are
present in the hot gas energised by the starburst processes, typically modelled with an optically thin, thermal spectrum with multiple temperature components (e.g., Moran et al. 1999; Jenkins et al. 2004; Warwick et al. 2007). 
Crucially, these components are detectable even in moderate quality \xmmn
data (e.g. Franceschini et al. 2003; Jenkins et al. 2004), providing a
test of whether the starburst dominates the X-ray emission.

The spectra were analysed using \xspec version 11.3.2 (Arnaud
1996).  For the \xmmn datasets we had three separate spectra for each
observation of each source (one from the pn camera, and two from the
MOS); these were analysed simultaneously, with a constant coefficient
inserted to model calibration uncertainties between the detectors.
The value of this coefficient was fixed to 1.0 for the pn, and allowed
to vary in the fits to the MOS data.  In practise, the value of this
constant varied by no more than $\pm 10\%$ between detectors in the
majority of observations.

As noted in section 3, most fits were performed using the standard
\xspec $\chi^2$ minimisation techniques.  However, some of the data
were so poor they did not meet the Gaussian criteria for $\chi^2$
fitting; consequentially we used the Cash statistics option in \xspec
to perform maximum likelihood fitting to these data. %Table~\ref{tab:stats} shows which data were analysed using $\chi^2$ statistics, and which data were analysed using Cash statistics. 
Cash statistics was only used on observations 7925, 9558, and 0112840101.  We began by analysing the background spectrum for each
source.  For the \chan data we constrained its form by fitting a
simple power-law continuum model to it, using the Cash statistics
option in \xspec (i.e., "cstat").  In the case of the \xmmn data in
observation 0112840101, we examined only the pn data as the datasets
were heavily background-dominated, and the MOS cameras only
contributed $\sim$ 20 - 30 source counts each.  The background data
had sufficient statistics to use $\chi^2$ fitting, and required a
broken power-law plus a narrow gaussian component to produce a good fit.
The source data itself was then analysed, but not background
subtracted.  Instead, the results of the fit to the background
spectrum were added as a fixed component to the spectral model, scaled
appropriately for the relative aperture sizes.  The model fitting was
then performed using Cash statistics.  The quality of the fit was
evaluated using the 'goodness' command, which we set to compare the
obtained C-statistic value to 1000 Monte-Carlo simulations based on
the best fitting model, and provides a percentage of simulated
datasets with a C-statistic value lower than that obtained in the fit.

The spectral fitting was conducted using a very basic absorbed
power-law continuum model.  The precise combination of models utilised
was TBABS*ZTBABS*ZPO in \xspec syntax, where the TBABS component
represents the foreground absorption column in our own Galaxy, the
ZTBABS is the absorption column local to the X-ray source (and hence
redshifted), and the ZPO is the redshifted power-law continuum. The
foreground Galactic absorption was set to a fixed value, calculated
using the NRAO dataset of Dickey \& Lockman (1990) in the \colden
Galactic hydrogen column density calculator within the \chan Proposal
Planning Toolkit\footnote{\tt
http://cxc.harvard.edu/toolkit/colden.jsp}, and shown for each galaxy
in Table~\ref{tab:tab0}.  The redshifts were derived from SDSS data,
and are also listed in Table~\ref{tab:tab0}. For both the Galactic
and intrinsic absorption columns, the abundance levels are set as per
Wilms, Allen and McCray (2000).  We show the result of these fits in
Table~\ref{tab:PLfits}.  The errors quoted in Table~\ref{tab:PLfits}
are the 90\% errors on the model parameters for one interesting
parameter, and the 68\% errors for the fluxes. As an example of the
data quality, we show the four spectra from observations of
J140052.5-014510 in Fig.~\ref{fig:fig1}.

All the $\chi^2$ fits provide statistically acceptable explanations of
the data with a simple absorbed power-law model (null hypothesis
probability $> 0.05$).  The goodness criteria for both fits of
J123719.3+114915 are also acceptable; however we note the Cash
statistics fit to the 7925 data for J082042.4+205715 has a poor
goodness (99\% of simulated power-law datasets providing a better Cash
statistic).  However, the data for this model is too poor to attempt
more complicated fits.

Table~\ref{tab:PLfits} shows that each of the sources displays a very
distinct spectrum. J082042.4+205715 has a relatively hard slope, with
$\Gamma \la 2$, but has little intrinsic absorption with a column $\la
10^{21}$ cm$^{-2}$.  J123719.3+114915 appears very hard with a
power-law slope $\Gamma < 1.6$, although its absorption column appears inconsistent
between the two observations. Finally, J140052.5-014510 has both a
very low absorption column of $\la 5 \times 10^{20}$ cm$^{-2}$, and
the softest spectrum with, $\Gamma \geq 2$. Note that all the spectral fits in Table~\ref{tab:PLfits} are shown in chronological order (see Table~\ref{tab:obs} for observation dates). 

Although the power-law models provide reasonable fits to the data, we
examined whether there was any evidence for further spectral
complexity by the addition of further components to the model.  We
only did this for the better datasets, i.e. those that allowed
$\chi^2$ statistics to be used (all observations of J140052.5-014510,
and the \xmmn observations of J082042.5+205715).  Given that starburst
spectra show strong collisionally-excited emission lines, we started
by adding a MEKAL component (Mewe et al. 1985, Mewe et al. 1986,
Kaastra, 1992) to each power-law model, setting each MEKAL component to the
appropriate redshift and to solar abundance.  In all cases, the improvements to the
fits were minimal. By setting the MEKAL temperature to 0.6 keV
(cf. Jenkins et al. 2004) and re-fitting for each dataset we were able
to place an upper limit on the contribution of a starburst-like MEKAL
component to each power-law dominated spectrum.  In the case of
J082042.5+205715 the 99\% upper limit was no more than 5.5\% of the
total 0.3 - 10 keV flux, whereas for J140052.5-014510 the 99\% upper
limit was just under 9\%.  Hence a hot gas starburst component can
only be present in the spectra of those two objects at a comparatively
minor level.

A separate additional component that may be present in the data is a
soft excess, as is found in many AGNs (e.g. Gierli{\'n}ski \& Done
2004).  We performed similar analyses to those for the addition of a
MEKAL, and found no strong evidence for the presence of a soft excess
in any dataset (maximum $\Delta\chi^2 < 8$ from the addition of a
DISKBB model with two additional degrees of freedom).

A final test we performed on the data was to constrain the presence of a
neutral Fe K line.  A high ($\ga 1$ keV) equivalent width
6.4 keV Fe K emission line is indicative of the presence of a
Compton-thick obscuring torus, blocking the direct line-of-sight to an
AGN (e.g. Awaki et al. 1991; Ghisellini, Haardt \& Matt 1994).  We
therefore added a narrow 6.4 keV Gaussian component to the spectra (at
the appropriate redshift) and calculated the 99\% upper limits on the
possible contribution of such a component to each spectrum.  We
tabulate the results in Table~\ref{tab:tabk}.  Clearly, the data for
J140052.5-014510 only rules out a $\sim 1$ keV equivalent width line
in one dataset (9557), and even that still has a rather high
equivalent width of $\sim 860$ eV allowed at the 99\% limit.  In
contrast, the 0505930301 data for J082042.5+205715 places a very
stringent 99\% upper limit of 140 eV on any possible Fe K line
contribution, clearly ruling out a high equivalent width line for this
object.

\begin{table}
\begin{center}
\caption{99\% upper limits on possible Fe K line equivalent widths}
\label{tab:tabk}
\begin{tabular}{cc}
\hline
Observation ID & Equivalent width limit \\
 & (keV) \\
\hline
{\bf 2XMMp J082042.5+205715}\\
0108860501 & $<$ 2.04 \\
0505930301 & $<$ 0.14 \\
{\bf 2XMMp J140052.5-014510}\\
0200430901 & $<$ 12.96 \\
0505930101 & $<$ 1.06 \\
0505930401 & $<$ 4.85 \\
9557 & $<$ 0.86 \\
\hline
\end{tabular}
\end{center}
\end{table}

\begin{figure}
\begin{center}
\vbox{
\psfig{figure=po1.ps,height=50mm,width=80mm,angle=270}
%\vspace={-0.0cm}
\psfig{figure=po2.ps,height=50mm,width=80mm,angle=270}
%\vspace={0.0cm}
\psfig{figure=po3.ps,height=50mm,width=80mm,angle=270}
%\vspace={0.0cm}
\psfig{figure=po4.ps,height=50mm,width=80mm,angle=270}
} 
\caption{The four observed spectra of the source 2XMMp
140052.5-014510, ordered from top to bottom by observation
date. The top three spectra are from {\it XMM-Newton\/} observations
(we only show EPIC-pn data for clarity), and the last spectrum is that
of the {\it Chandra\/} ACIS-S 9557 observation. The four spectra are
plotted against the absorbed power-law model best fitting to the {\it Chandra\/} data to
illustrate changes in the spectrum between different observing
epochs. $\Delta\chi$ is plotted underneath the spectra.}
\label{fig:fig1}
\end{center}
\end{figure}

\begin{figure*}
\begin{center}
\hbox{\psfig{figure=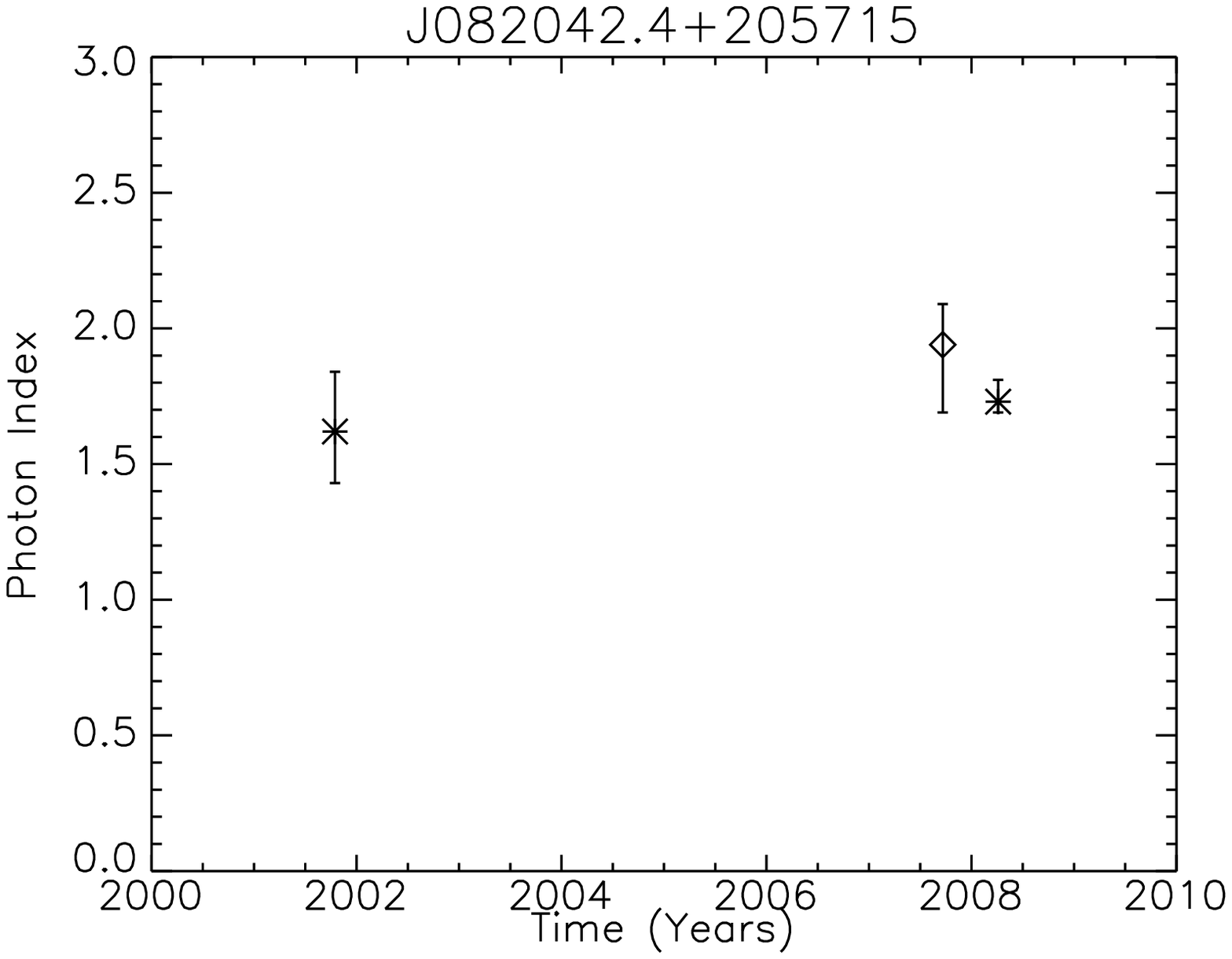,height=5cm,width=6cm}
\hspace{-0.3cm}
\psfig{figure=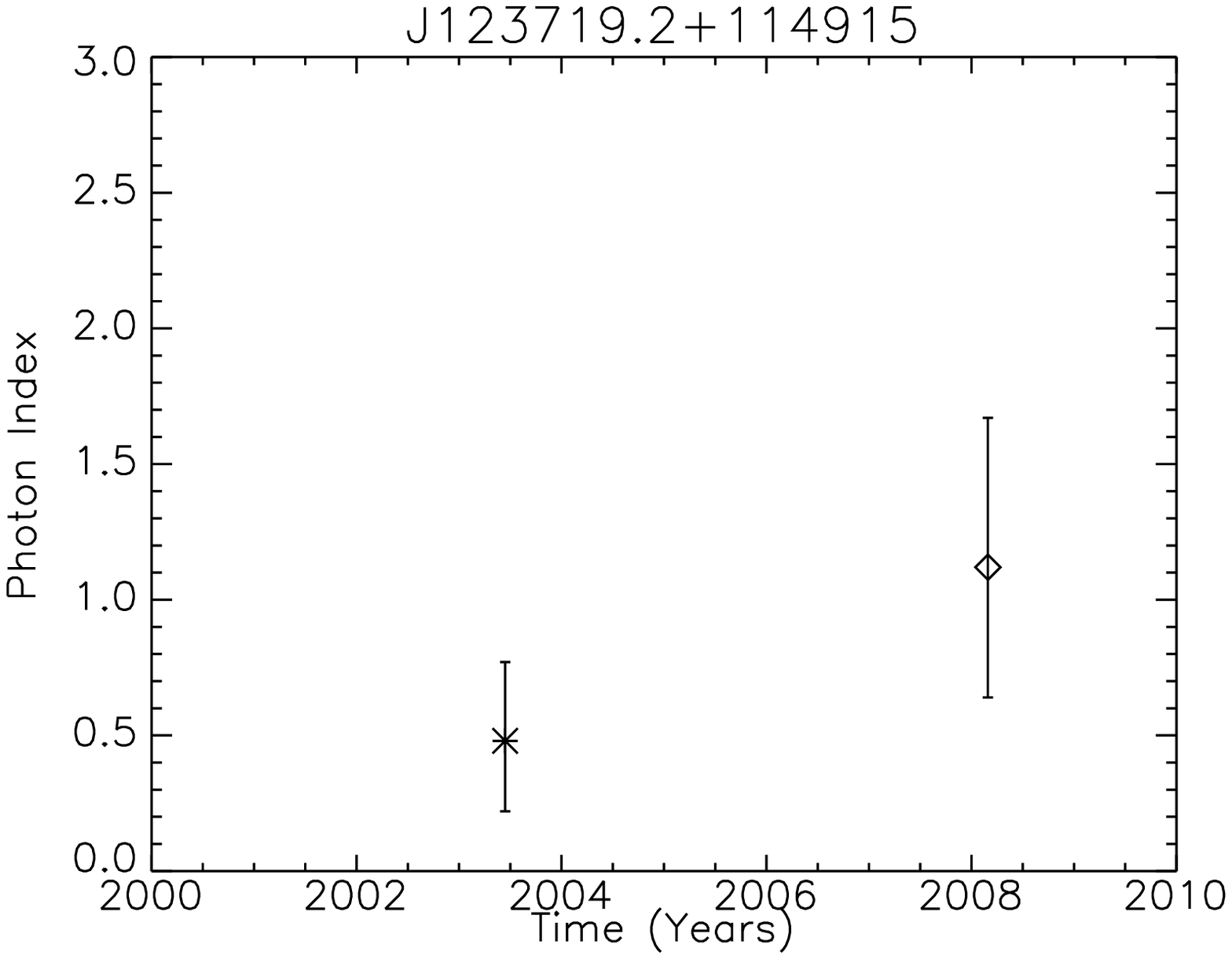,height=5cm,width=6cm}
\hspace{-0.3cm}
\psfig{figure=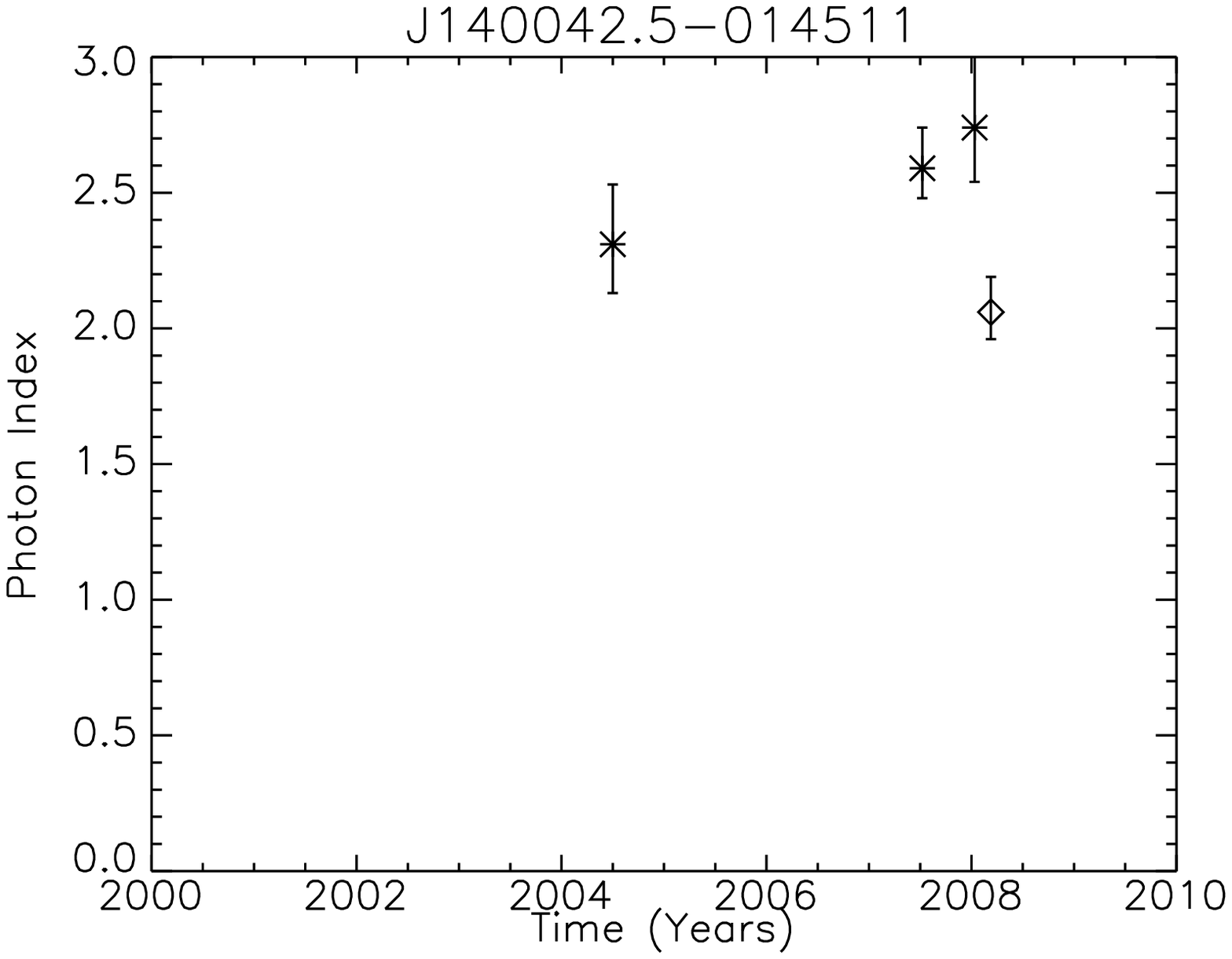,height=5cm,width=6cm}}
\vspace{0.0cm}
\hbox
{\psfig{figure=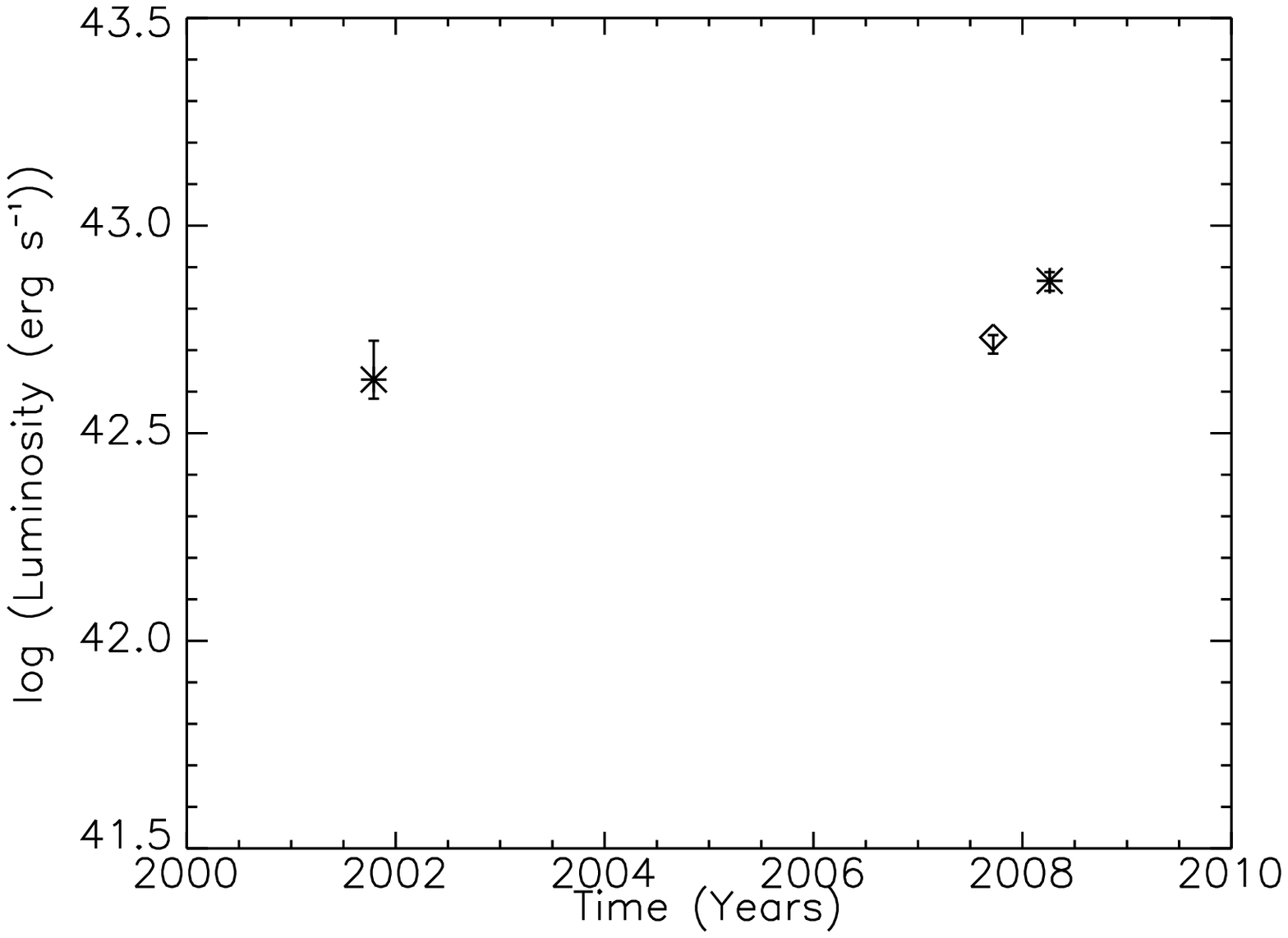,height=5cm,width=6cm}
\hspace{-0.3cm}
\psfig{figure=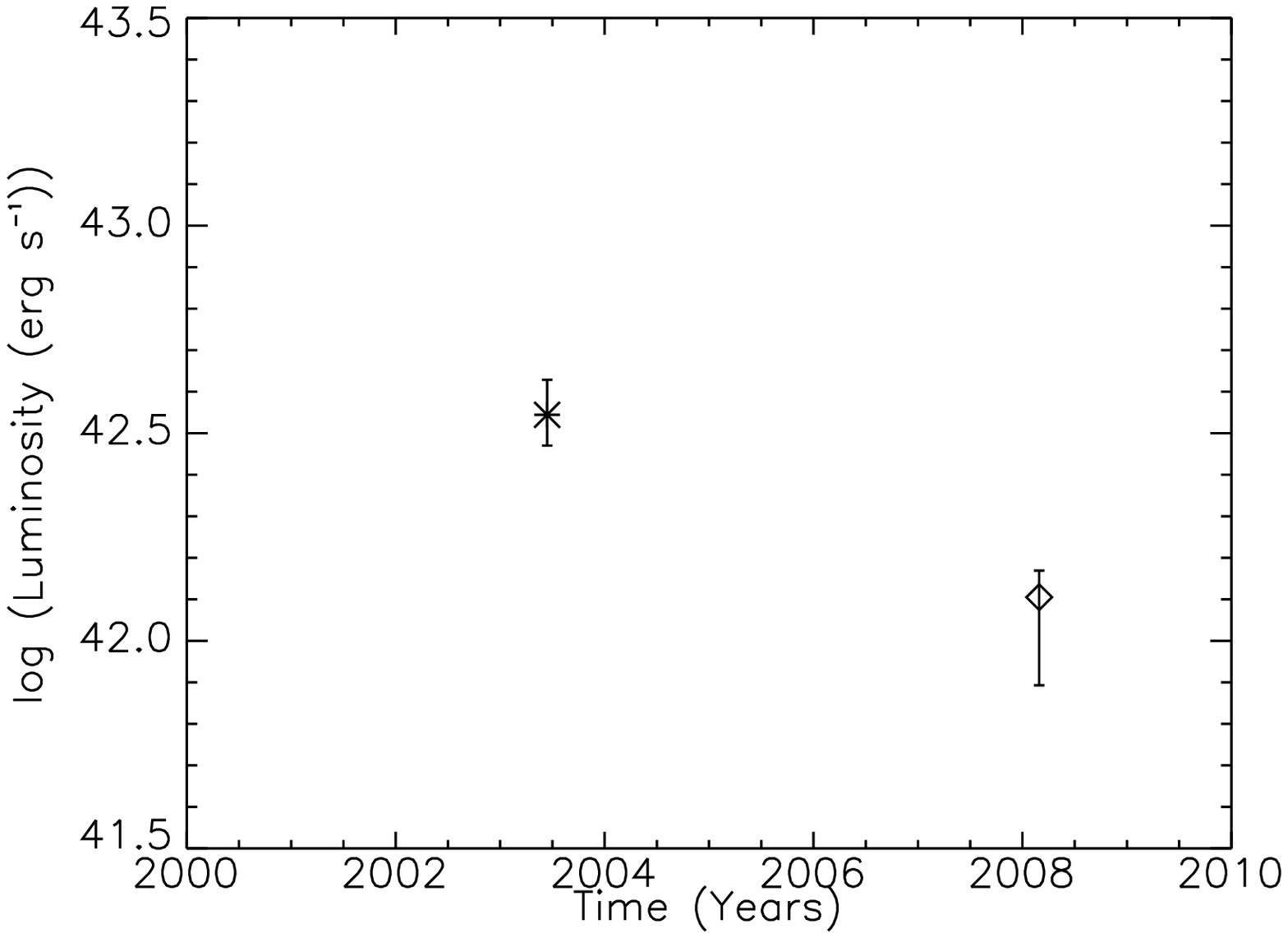,height=5cm,width=6cm}
\hspace{-0.3cm}
\psfig{figure=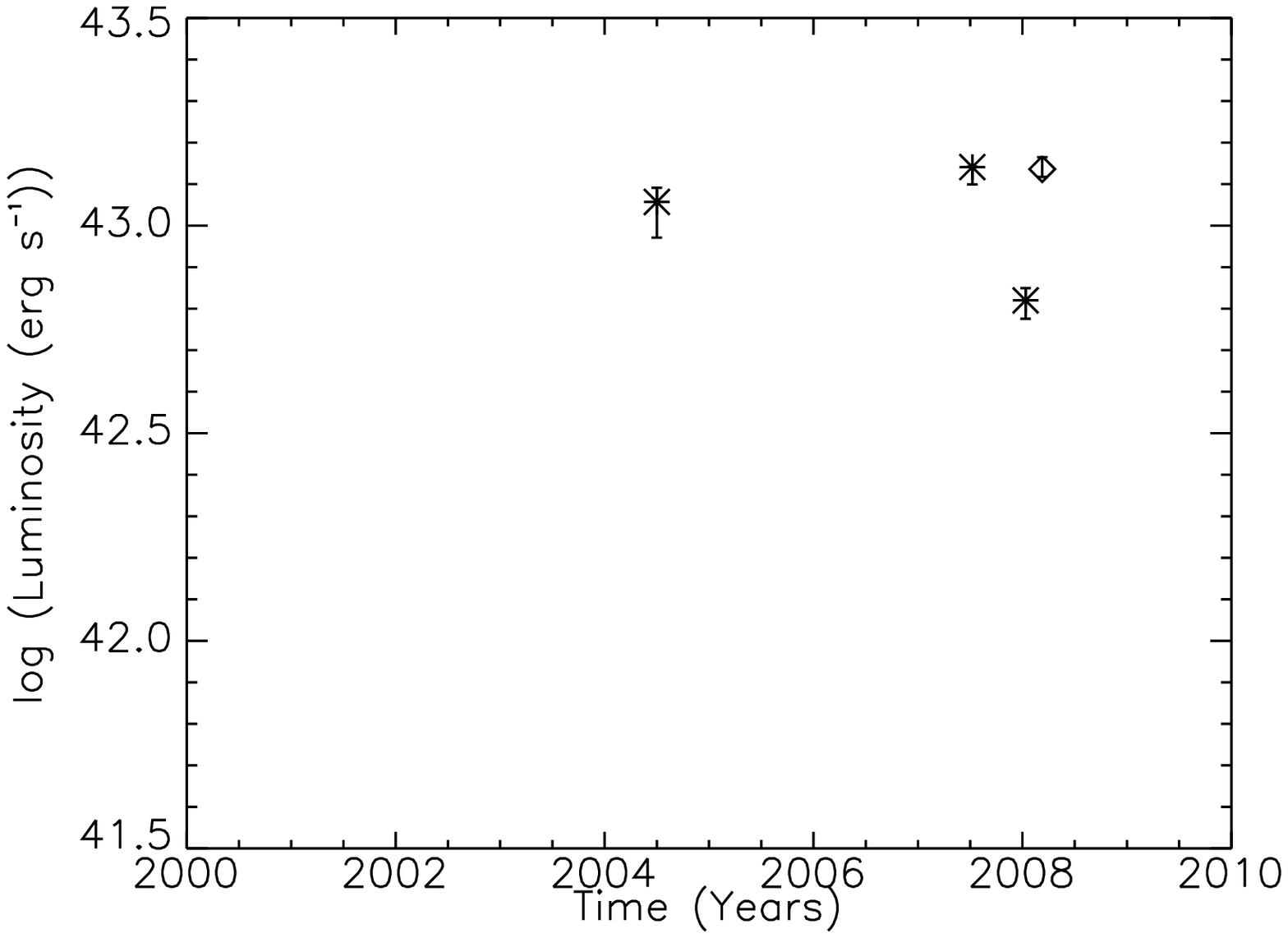,height=5cm,width=6cm}}
\caption{Source luminosity and photon index variability in the 0.3 to
10.0 keV X-ray band.  The columns show data from (left -- right)
J082042.4+205715, J123719.3+114915 and J140052.5-014510 respectively. The errors shown are calculated within a 90$\%$ Gaussian range.
{\it Top row:\/} Measured photon index as a function of time.  {\it
Bottom row:\/} Observed luminosity as a function of time. The errors shown are calculated within a 68$\%$ Gaussian range. The values
are taken from the spectral fits presented in Section 4.3 or
calculated as per the text, and each row is plotted on the same scale
to permit direct comparison.  \xmmn data points are indicated by
asterisks, and \chan data by diamonds.}
\label{fig:fig4}
\end{center}
\end{figure*}

\subsection{Long term and spectral variability}

Our final set of diagnostics are taken from studying possible
variations of both the X-ray flux and X-ray spectral shape between the
different observation epochs, which are separated by gaps of months to
years.  Again, we would expect to see little variation from
spatially extended X-ray emission regions composed of many components
such as a starburst region, whereas AGN can vary both temporally and
spectrally over such timescales (e.g. Taylor, Uttley \& McHardy 2003;
Sobolewska \& Papadakis 2009).

Within the spectral data analysed above we do see evidence for
spectral variability between different observation epochs.  We
illustrate this in Fig.~\ref{fig:fig1} in which we present the four
different spectra of J140052.5-014510 with residuals (pn only in the three \xmmn
datasets, for the sake of clarity), and compare each spectrum to the
model best fitting to the Obs ID 9557 \chan data.  Clearly the spectra
in the top three panels deviate from the shape of the last spectrum,
demonstrating some degree of variability in the data.  We investigate
whether this is due to strong variability in the underlying spectral
shape in the top three panels of Fig.~\ref{fig:fig4}, where we compare
the measured values of the photon indices between the different
observations.  It is clear that the best fitting values indicate that
the spectrum of J140052.5-014510 does vary significantly, over periods
as short as months.  It is less clear whether this is the case for the
other objects; their power-law photon indices are at least consistent
at the 2$\sigma$ level (or better) over their observations.

The long-term luminosity variations are presented as the bottom three
panels, where the luminosity is calculated from the fluxes in
Table~\ref{tab:PLfits}\footnote{These are the observed 0.3 - 10 keV luminosities, not
the rest frame values.  However, we note that the similar, low
redshifts of the three objects lead to relatively small corrections
when converting to rest frame luminosities, with the increases of
$\sim 10\%$ for the hard spectrum source J123719.3+114915, and $< 5\%$
for the other two objects.}. All three objects clearly demonstrate
significant luminosity variations over timescales of months,
consistent with relatively compact objects.

\section{Discussion}

The basic question posed at the beginning of this work was: what can power more than $10^{42}$ erg s$^{-1}$ of X-ray luminosity from a
galaxy with no unambiguous signatures of AGN activity? We have examined this by
selecting three galaxies at a redshift $\sim$ 0.1 from SDSS for further study at X-ray
energies, based on the requisite high apparent X-ray luminosity and an
optical spectrum betraying only signs of star formation processes.
The work presented in the previous section details the results of
several simple tests with the power to diagnose whether the X-ray
emission originates in star formation processes, or an AGN.

We find that all three sources appear more consistent with AGNs than
starbursts.  Firstly, all are point-like in the X-ray images
(cf. Fig.~\ref{fig:figrad}; Table~\ref{tab:tabpsf}), although this is
not convincing evidence in isolation given that the physical size
limits on any extended component are of the order $\la 1$ kpc at the
distance of the galaxies, permitting at least compact nuclear
starburst emission to be present.  Far more convincing evidence comes
from the time variability characteristics of the X-ray emission, with
flux variations of factors $\ga 2$ seen for all three objects
(Fig.~\ref{fig:fig4}; Table~\ref{tab:PLfits}) implying that at least
50\% of the X-ray luminosity originates in a region no more than a few
light-years across.  In the specific case of J140052.5-014510, we see
its flux halving and then returning to its original value in a period of
6 hours in the \chan ACIS-S light curve, implying the X-ray emission
originates from a solar system-sized object, which -- for a luminosity
around $10^{43}$ erg s$^{-1}$ -- has to be due to an AGN.  Finally, the X-ray
spectra of the three objects are all well-described by a power-law
continuum, consistent with moderate quality AGN data and not the
strong emission line-dominated spectra seen from starbursts.  It
should be noted that this spectral deconvolution implies dominance by
an AGN-like component, but does not rule out a bright X-ray starburst.
In fact, the 99\% upper limits on the MEKAL thermal plasma
contributions to each spectrum still allow for in the region of $\sim
2 \times 10^{41}$ erg s$^{-1}$ of starburst-heated gas in
J082042.4+205715 (comparable to the thermal component seen in many
ULIRGs, e.g. Franceschini et al. 2003), and up to $\sim 10^{42}$ erg
s$^{-1}$ in J140052.5-014510.

An interesting alternative to an AGN that could plausibly produce extreme
X-ray luminosities without an AGN-like optical signature is a
hyperluminous X-ray source (HLX, e.g. Gao et al. 2003), the brightest of
the off-nuclear ultraluminous X-ray sources (ULXs, see Roberts et al.
2007; Feng \& Soria 2011 for reviews).  The most luminous of these
objects, ESO 243-49 HLX-1, is as bright as $10^{42}$ erg s$^{-1}$ in 0.3
-- 10 keV X-rays at peak (Farrell et al. 2009; Servillat et al. 2011), and
so if a similar object were located close to the nucleus of a moderate
redshift galaxy it might indeed share the observational characteristics of
the three objects in this work.  However, such objects are rare compared
to bona fide AGNs -- only a handful of ULXs with $L_{\rm X} > 5 \times
10^{40} \rm ~erg~s^{-1}$ are detected within 100 Mpc in the 2XMM-DR1
catalogue (Walton et al. 2011), far fewer in number than AGNs in a similar
volume (e.g. Ho 2008), and with the exception of ESO 243-49 HLX-1 (which
may itself be an accreted dwarf galaxy nucleus, e.g. Soria et al. 2010)
they are far less luminous than galactic nuclei.  We therefore do not
regard this as a likely nature for any of our objects.

So if these objects are all AGN-dominated in the X-rays, can we say
anything about the type of AGN from the X-ray data? Firstly, only one
of the objects -- J123719.3+114915 -- appears hard ($\Gamma$ $<$ 1.6) and perhaps
relatively highly absorbed ($N_{\rm H}$ $\sim$ 10$^{22}$ cm$^{-2}$ during the observation 9558), consistent with a classic type 2 object,
although the poor quality of this data should be noted. The other two
objects have much lower columns, with both constrained at $\la
10^{21}$ cm$^{-2}$, more consistent with type-1 AGNs. This is
supported when the data are sufficiently good to place constraints
on the presence of a neutral Fe K line; in both cases high equivalent
width ($> 1$ keV) lines are ruled out, rejecting a strong reflection
component originating in the torus of a type-2 AGN (e.g. Awaki et
al. 1991).  However, both have different spectral slopes, with
J140052.5-014510 markedly softer than J082042.4+205715. While the
slope of J082042.4+205715 ($\Gamma$ $\la$ 2) is fairly typical for type-1 AGNs
(e.g. Nandra \& Pounds 1994), the soft slope of J140052.5-014510 ($\Gamma$ $\ge$ 2) is
more unusual, and consistent with slopes seen in some narrow-line
Seyfert 1 galaxies (NLS1s; Vaughan et al. 1999).  However, there is no
strong evidence of a distinct soft excess -- as is seen in NLS1s -- in
either object. A piece of optical evidence indicating the possible presence of NLS1 activity is the line width of H$\beta$ emission. By definition, NLS1 galaxies emit H$\beta$ with line widths $<$ 2000 km s$^{-1}$ (Wang \& Zhang 2007). We found that the three sources emit H$\beta$ at widths considerably lower than this upper limit ($<$ 300 km s$^{-1}$). However, these line widths are also consistent with those of star forming galaxies.   

%NLR obscuration
%BLR obscuration + low edd_lim
%

So why is there no signature of AGN activity in these objects' optical spectra? A number of reasons for this have been suggested in previous work focused on the nature of so-called 'elusive AGNs', i.e., sources that show no signs of AGN activity in the optical regime, but display signs of AGN activity in the X-ray band. One possibility is that the regions within the galaxy producing narrow and broad line emission typically seen in AGNs have been rendered 'invisible' by obscuration. For example, obscuration of AGN activity within NGC 4945 by a heavy dusty torus was deduced by Iwasawa et al (1993). Obscuration of the narrow line region may also be caused by spherical gas clouds shrouding the nuclei of galaxies, such as PKS 031208 (Comastri et al. 2002). However, obscuration is not necessarily caused by gas and dust close to the galactic nucleus. Rigby et al. (2006) reported that from a sample of 31 AGNs at redshifts of 0.5 $<$ $z$ $<$ 0.8, only the most face-on galaxies displayed optical AGN activity. This implies that the obscuration of both the broad and narrow line regions may well be caused by host galaxy dust along our line of sight (e.g., Moran et al. 1996; Rigby et al. 2006; Goulding $\&$ Alexander 2009). However, given that the X-ray spectral analysis of the three sources yielded hydrogen column density upper limit values $\leq$ 10$^{21}$ cm$^{-2}$, and that the calculated extinction values from the optical spectra for the galaxies are 1.11 mag $<$ A$_v$ $<$ 2.86 mag, obscuration is unlikely to be the cause for the elusiveness of optical AGN signatures in the sources. 

Another possibility is that the broad line regions (BLRs) that should be signatures of AGN activity may not be seen because of low accretion rates. According to Nicastro et al. (2000), AGN BLRs may form in disc winds at a critical distance within the accretion disc. This formation occurs at an accretion rate $>$ 10$^{-3}$ $\dot{m}$ ($=~\dot{M}/\dot{M}_{\rm Edd}$). If the accretion rate is lower than this, the BLR cannot form. A recent study by Trump et al. (2011) deduced that the BLR of an AGN becomes invisible if the accretion rate drops below 10$^{-2}$ $\dot{m}$. This is because at such low accretion rates, radiatively inefficient accretion flows (RIAFs, Yuan $\&$ Narayan 2004) occur and expand close to the inner radius of the accretion disc. RIAFs lack the optical/UV emission of an optically thick accretion disc and therefore cannot ionise BLRs (Yuan $\&$ Narayan 2004; Trump et al. 2011).  
%The low luminosity of the AGN within Q2131-427 my well explain the absence of broad lines in its optical spectrum (Panessa et al. 2009).  
In order for a source to have an accretion rate lower than 10$^{-2}$ $\dot{m}$ and to be displaying a luminosity of 10$^{42}$ erg s$^{-1}$, the mass of the accreting black holes would have to be $\ge$ 8 $\times$ 10$^{7}$ M$_{\sun}$. Using the M-$\sigma$ relation (Gultekin. et al. 2009) with emission values from the SDSS DR7 archive, we estimated the masses of the black holes in the centre of the sources to be 5 $\times$ 10$^{8}$ M$_{\sun}$, 3 $\times$ 10$^{7}$ M$_{\sun}$, and 2 $\times$ 10$^{8}$ M$_{\sun}$ for J082042.4+205715, J123719.3+114915, and J140052.5-014511, respectively. Therefore, low levels of accretion activity is a possible explanation for the lack of the BLRs in two of the three galaxies' nuclei.

One more possibility is that the AGN activity has been optically
diluted by starlight, and indeed star formation signatures, within the host galaxy. Caccianiga et al (2007) found that from a sample of of 9
star forming optically elusive galaxies, all of their AGN signatures were diluted by galactic starlight. A
similar result was seen in the study of optically elusive AGNs by Trump
et al (2009). In this study, 70\% of the 48 optically elusive AGN
were diluted by galactic starlight. 
If we investigate the emission-line diagnostic diagrams shown in Fig.~\ref{fig:bpt}, we can see that all three sources' flux ratios are positioned, in both cases, well below the Kewley et al. (2001) maximum starburst line (it is worth reiterating that the Transition region of the [N{\small II}]/H$\alpha$ diagnostic diagram does contain the flux ratios of pure star forming galaxies). 
%in the [S{\small II}]/H$\alpha$ 
%star forming region. % and the [N{\small II}]/H$\alpha$ Transition region (it is worth reiterating that the Transition region of the [N{\small II}]/H$\alpha$ diagnostic diagram does contain flux ratios of pure star forming galaxies). 
It is likely that star formation in the galaxies has strengthened the H$\alpha$ and H$\beta$ emission fluxes, thus diminishing the [S{\small II}]/H$\alpha$, [N{\small II}]/H$\alpha$, and [O{\small III}]/H$\beta$ values. If we remove the galactic star formation signatures, our objects would likely end up within the AGN area of the diagram. Thus, we conclude that the elusiveness of optical AGN signatures within our galaxies is most likely due to dilution by stellar activity.

Our results are similar to those of Jia et al. (2011), where they investigate X-ray, optical, and infra-red properties of 6 Lyman break analogue (LBA) galaxies with %[N{\small II}]/H$\alpha$ and [O{\small III}]/H$\beta$ 
optical line ratios similar those of our sources. They deduced that a combination of intense star formation and AGN activity is likely present within the LBAs by studying their multi-wavelength properties. Unlike the work of Jia et al., we were able to deduce AGN presence by looking at X-ray data alone, due to the superior quality of our X-ray data. Nevertheless, both our results and those of Jia et al. agree on the presence of AGN activity and star formation within both galaxy samples.

\section{Conclusion}

Are galaxies that are both X-ray bright ($L_{\rm X}$ $>$ 10$^{42}$ erg s$^{-1}$) and display no clear optical signatures of AGN activity truly powered by star formation? Or are these sources powered by an optically-hidden AGN? In an attempt to answer these questions, we have investigated the X-ray properties of 3 sources that were taken from a cross correlation of the SDSS DR5 and 2XMMp DR0 catalogues, that have both no optical signatures of AGN activity, X-ray luminosities exceeding 10$^{42}$ erg s$^{-1}$, and, crucially, relatively good X-ray data (a few hundred to few thousands of counts). This investigation was performed using data from 9 observations of the three sources, taken using both \xmmn and \chan.

To test for the presence of activity within these sources' nuclei, we adopted 4 tests: point spread analysis, intra-observational variability analysis, spectral analysis, and long term spectral and timing variability analysis. Investigating the point spread function of the 3 sources indicates that the sources' X-ray emission in not extended. The compact nature of the sources is also evident in the long-term photon count rate variability of the three sources. J140052.5-014511 displays short term count rate variability (during one \chan observation), and long-term variability in its spectral shape, emphasising this source in particular must be very compact. All of this evidence points towards the three sources being X-ray powered by point-like objects.

The spectral analysis of the sources indicates the lack of star formation emission from the three sources in the X-ray band. All three sources' spectra are well fitted with a simple absorbed power-law model. When fit with an absorbed power-law plus MEKAL component model, $>$ 90$\%$ of the sources' flux is constrained to be derived from the power-law component. Note, however, that star formation may still be causing a very small proportion of X-ray emission within the nuclei of the galaxies.  
%In particular, a 99$\%$ flux upper limit of up to $\sim$ 2 $\times$ 10$^{41}$ erg s$^{-1}$ is seen to be coming from the star forming component when the spectra of J082042.4+205715 is fit with the power-law plus MEKAl model. $\sim$ 10$^{42}$ erg s$^{-1}$ of luminosity is seen as a 99$\%$ upper limit to the MEKAL component flux of the J140052.5-014511 spectrum when fit with the same model. Both of these fluxes still imply the presence of a relatively bright star forming region within the galaxies' nuclei. As well as the X-ray evidence for the presence of star formation, evidence can be seen in the optical spectra of the sources. 

The results yielded from the X-ray studies of the sources infers the presence of AGN activity within the galaxies' nuclei. But why are these AGNs optically elusive? A number of possibilities have been suggested as to the cause of this optical elusiveness. The most likely cause for this elusiveness is that the AGNs are being optically diluted by stellar emission from within their host galaxies. 

At the beginning of this study, we asked whether galaxies with X-ray luminosities $>$ 10$^{42}$ erg s$^{-1}$ and that display no signs of AGN activity in the optical band are truly powered by star formation, or are they powered by a hidden AGN. By studying the X-ray properties of three medium redshift objects that display these characteristics, we have discovered that such galaxies are indeed powered by AGN activity. Therefore, we have deduced that the common assumption that galaxies which display luminosities $>$ 10$^{42}$ erg s$^{-1}$ host AGNs appears to be safe.

\section*{Acknowledgements}

This work is based on observations obtained with XMM-Newton, an ESA science 
 mission with instruments and contributions directly funded by ESA Member 
 States and the USA (NASA). This research has made use of software provided by the Chandra X-ray Center (CXC) in the application package CIAO. Funding for the SDSS and SDSS-II has been provided by the Alfred P. Sloan Foundation, the Participating Institutions, the National Science Foundation, the U.S. Department of Energy, the National Aeronautics and Space Administration, the Japanese Monbukagakusho, the Max Planck Society, and the Higher Education Funding Council for England. The SDSS Web Site is http://www.sdss.org/.    The SDSS is managed by the Astrophysical Research Consortium for the Participating Institutions. This research has made use of the NASA/IPAC Extragalactic Database (NED) which is operated by the Jet Propulsion Laboratory, California Institute of Technology, under contract with the National Aeronautics and Space Administration. This research has made use of the SIMBAD database, operated at CDS, Strasbourg, France.\\

\label{lastpage}

{\bf References} 

\noindent\hangindent=0.5cm\hangafter=1
Alexander D. M., et al. 2003, AJ, 126, 539

\noindent\hangindent=0.5cm\hangafter=1 
Alexander D. M., Bauer F. E., Chapman S. C., Smail I., Blain A. W,. Brand, W. N., Iviso, R. J. 2005, ApJ, 632, 736

\noindent\hangindent=0.5cm\hangafter=1 
Arnaud K. A. 1996, Astronomical Data Analysis Software and Systems V, eds. Jacoby G. and Barnes J., p17, ASP Conf. Series volume 101

%\noindent\hangindent=0.5cm\hangafter=1 
%Awaka, H. 1997, ASP conf. series Vol 113m p. 44

\noindent\hangindent=0.5cm\hangafter=1 
Awaki H., Koyama K., Inoue H., Halpern J. 1991, PASJ, 43, 195

%\noindent\hangindent=0.5cm\hangafter=1 
%Barth, A. J., Ho, L. C., Filippenko, A. V., Sargent, W. L. W. 1998, ApJ, 49, 133B

\noindent\hangindent=0.5cm\hangafter=1
Barger A. J., Cowie L. L., Brandt W. N., Capak P., Garmier G. P., Hornschemeier A. E., Steffan A. T., Wehner E. H. 2002, AJ, 124, 1839

\noindent\hangindent=0.5cm\hangafter=1
Barger A. J. 2003, AIPC, 666, 205

\noindent\hangindent=0.5cm\hangafter=1
Bauer F. E., et al. 2004, AdSpR, 34, 2555

%\noindent\hangindent=0.5cm\hangafter=1 
%Berghea, C. T., Weaver, K. A,. Colbert, E. J. M., Roberts, T. P. 2008, ApJ, 687, 471

\noindent\hangindent=0.5cm\hangafter=1 
Blustin A. J., Page M. J., Fuerst S. V., Branduardi-Raymont G., Ashton C. E. 2005, A\&A, 431, 111

\noindent\hangindent=0.5cm\hangafter=1
Brandt W. N. Hasinger G. 2005, ARA$\&$A, 43, 827

\noindent\hangindent=0.5cm\hangafter=1
Bressan A., Silvia L., Granato G. L. 2002, A$\&$A, 392, 377  
%\noindent\hangindent=0.5cm\hangafter=1 
%Brock, D., Joy, M., Lester, D. F., et al. 1988, ApJ, 329, 208

\noindent\hangindent=0.5cm\hangafter=1 
Caccianiga A., Severgnini P., Della Ceca R., Maccacaro T., Carrera F. J., Page M. 2007, A$\&$A, 470, 557

\noindent\hangindent=0.5cm\hangafter=1 
Cash W. 1979, ApJ, 228, 939

\noindent\hangindent=0.5cm\hangafter=1 
Ciliegi P., Elvis M., Wilkes B. J., Boyle B. J., McMahon R.G. 1997, MNRAS, 284, 401

%\noindent\hangindent=0.5cm\hangafter=1 
%Civano, F., Mignoli, M., Comastri, A., et al. 2007, A$\&$A, 476, 1223

%\noindent\hangindent=0.5cm\hangafter=1 
%Colbert, E. J. M., Mushotzky, R. F. 1999, ApJ, 519, 89

\noindent\hangindent=0.5cm\hangafter=1 
Colless M., et al. 2001, MNRAS, 328, 1039

\noindent\hangindent=0.5cm\hangafter=1 
Comastri A., et al. 2002, ApJ, 571, 771

%\noindent\hangindent=0.5cm\hangafter=1 
%Dewangan, G. C., Titarchuk, L., Griffiths, R. E. 2006, ApJ, 637, 21

%\noindent\hangindent=0.5cm\hangafter=1 
%Dewangan, G. C., Misra, R., Rao, A. R., Griffiths, R. E.  2009, arXiv, 0905, 3819D

\noindent\hangindent=0.5cm\hangafter=1 
Dickey J. M., Lockman F. J. 1990, ARA\&A, 28, 215

\noindent\hangindent=0.5cm\hangafter=1
Eckart M. E., Stern D., Helfand D. J., Harrison F. A., Mao P. H., Yost S. H. 2006, ApJS, 165, 19

%\noindent\hangindent=0.5cm\hangafter=1 
%Erracleous, M., Shields, J. C., Chartas, J., Moran, E. C. 2002, ApJ, 565, 108

%\noindent\hangindent=0.5cm\hangafter=1 
%Fabianno, G. 1988, ApJ, 330, 672F

\noindent\hangindent=0.5cm\hangafter=1
Elvis M., Civano F., Vignali C., et al. 2009, ApJS, 184, 158 

\noindent\hangindent=0.5cm\hangafter=1
Farrell S., Webb N. A., Barret D., Godet O., Rodriguies J. M. 2009, Natur, 460, 73  

\noindent\hangindent=0.5cm\hangafter=1
Feng H., Soria R. 2011, arXiv, 1109, 1610

\noindent\hangindent=0.5cm\hangafter=1 
Franceschini A., et al. 2003, MNRAS, 343, 1181

\noindent\hangindent=0.5cm\hangafter=1
Gao Y., Wang D. Q., Appleton P. N., Lucas R. A. 2003, ApJ, 596, 171

%\noindent\hangindent=0.5cm\hangafter=1 
%Fried, J. W., Schulz, H. 1983, A$\&$A, 118, 166

%\noindent\hangindent=0.5cm\hangafter=1 
%Genzel, R., Lutz, D., Sturm, E., et al. 1998, A$\&$A, 498, 579

\noindent\hangindent=0.5cm\hangafter=1 
Ghisellini G., Haardt F., Matt G. 1994, MNRAS, 267, 743

\noindent\hangindent=0.5cm\hangafter=1
Giacconi R., et al. 2001, ApJ, 551, 624

\noindent\hangindent=0.5cm\hangafter=1 
Gierli{\'n}ski M. Done C. 2004, MNRAS, 349, 7

\noindent\hangindent=0.5cm\hangafter=1
Goulding A. D., Alexander D. M. 2009, ASPC, 408, 59

\noindent\hangindent=0.5cm\hangafter=1
Goulding A. D., Alexander D. M. 2009, MNRAS, 398, 1165

\noindent\hangindent=0.5cm\hangafter=1
Gultekin K., et al. 2009, ApJ, 698, 198

\noindent\hangindent=0.5cm\hangafter=1
Grimm H. -J., Gilfanov M., Sunyaev R. 2003, MNRAS, 339, 793 

%\noindent\hangindent=0.5cm\hangafter=1 
%Gladstone, J., Roberts, T. P., Done, C. 2009, MNRAS, 397, 1836G

%\noindent\hangindent=0.5cm\hangafter=1 
%Goncalvez, A. C., Veron-Cetty, M. P., Veron, P. 1999, A$\&$AS, 135, 437

%\noindent\hangindent=0.5cm\hangafter=1 
%Gonzalez Deldago, R. 2002, ASPC, 258, 101

%\noindent\hangindent=0.5cm\hangafter=1 
%Gonzal\'{e}z-Mart\'{i}n, O., Masegosa, J., M\'{a}rquez, I., Guerrero, M. A., Dultzin-Hacyan, D. 2006, A$\&$A 460, 45

%\noindent\hangindent=0.5cm\hangafter=1 
%Heil, L. M., Vaughan, S., Roberts, T. P. 2009, MNRAS, 397, 1061H

\noindent\hangindent=0.5cm\hangafter=1
Hickox R. C., Markevitch M. 2006, ApJ, 645, 95

%\noindent\hangindent=0.5cm\hangafter=1 
%Ho, L. C. 1999, Adv. Space. Res., 23, 813

\noindent\hangindent=0.5cm\hangafter=1
Ho L. C. 2008, ARA$\&$A, 46, 475

\noindent\hangindent=0.5cm\hangafter=1
Hornschemeier et al. 2001, ApJ, 554, 742

\noindent\hangindent=0.5cm\hangafter=1
Hornschemeier A. E., Heckman T. M., Ptak A. F., Tremonti C. A., Colbert E. J. M. 2005, AJ, 129, 86

\noindent\hangindent=0.5cm\hangafter=1 
Iwasawa K., Koyama K,. Awaki H., Kunieda H., Makishima K., Tsuru T., Ohashi T., Nakai N. 1993, ApJ, 409, 155

%\noindent\hangindent=0.5cm\hangafter=1 
%Iwasawa, K., Comastri, A. 1998, MNRAS, 297, 1219
 
\noindent\hangindent=0.5cm\hangafter=1 
Jenkins L. P., Roberts T. P., Ward M. J., Zezas A. 2004, MNRAS, 352, 1335

\noindent\hangindent=0.5cm\hangafter=1 
Jia J., Ptak A., Heckamn T. M., Overzier R. A., Hornschemeier A., LaMassa S. M. 2011, ApJ, 731, 55

\noindent\hangindent=0.5cm\hangafter=1 
Kaastra J. S. 1992, An X-Ray Spectral Code for Optically Thin Plasmas (Internal SRON-Leiden Report, updated version 2.0)

\noindent\hangindent=0.5cm\hangafter=1
Kauffmann G., et al. 2003, MNRAS, 346, 1055

\noindent\hangindent=0.5cm\hangafter=1
Kewley L. J., Dopita M. A., Sutherland R. S., Heisler C. A., Trevema J. 2001, ApJ, 556, 121

\noindent\hangindent=0.5cm\hangafter=1
Kim M., et al. 2007, ApJS, 169, 401

%\noindent\hangindent=0.5cm\hangafter=1 
%K\"ording, E., Falcke, H., Markoff, S. 2002, A$\&$A, 382, 13

%\noindent\hangindent=0.5cm\hangafter=1 
%Lamer G., McHardy I.M., Uttley P., Jahoda K. 2003, MNRAS, 338, 323

\noindent\hangindent=0.5cm\hangafter=1 
Lawrence A., Watson M. G., Pounds K. A., Elvis M. 1987, Nature, 325, 694

\noindent\hangindent=0.5cm\hangafter=1
Lehmer B. D., et al. 2005, ApJS, 161, 21

\noindent\hangindent=0.5cm\hangafter=1
Lehmer B. D., et al. 2008. ApJ, 681, 1163 

\noindent\hangindent=0.5cm\hangafter=1
Lira P., Ward M. J., Zezas A., Alonso-Herrero A., Ueno, S. 2002, MNRAS, 330, 259

\noindent\hangindent=0.5cm\hangafter=1
Lumb D. H., Warwick R. S., Page M., De Luca A. 2002, A$\&$A, 398, 93

%\noindent\hangindent=0.5cm\hangafter=1 
%Magorrian, J., Tremaine, S., Richstone, D., et al. 1998, AJ, 115, 2285

\noindent\hangindent=0.5cm\hangafter=1 
Maiolino R., et al. 2003, MNRAS, 344, 59

%\noindent\hangindent=0.5cm\hangafter=1 
%Makishima, K., Maejima, Mitsuda, K., et al. 1986, ApJ, 308, 635

%\noindent\hangindent=0.5cm\hangafter=1 
%Mao, Y., Wang, J., Wei, J. 2009, arXiv, 0904, 4328M

\noindent\hangindent=0.5cm\hangafter=1 
Markowitz A., et al. 2003, ApJ, 593, 96

\noindent\hangindent=0.5cm\hangafter=1 
Marconi A., Oliva E., van der Werf P. P., Maiolino R., Schreier E. J., Macchetto F., Moorwood A. F. M. 2000, A$\&$A, 357,24  

\noindent\hangindent=0.5cm\hangafter=1 
Mewe R., Gronenschild E. H. B. M., van den Oord G. H. J. 1985, A$\&$AS, 62, 197

\noindent\hangindent=0.5cm\hangafter=1
Mewe R., Lemen J. R., van den Oord G. H. J. 1986, A$\&$AS, 65, 551M

%\noindent\hangindent=0.5cm\hangafter=1
%Mitsuda, K., Matsouka, M., Inoue, H., et al. 1984, PASJ, 36, 741M

\noindent\hangindent=0.5cm\hangafter=1
Moran E. C., Lehnert M. D., Helfand D. J, 1999, ApJ, 526, 649 

\noindent\hangindent=0.5cm\hangafter=1
Moran E. C., Halpern J. P., Helfand D.J. 1996, ApJS, 106, 341

\noindent\hangindent=0.5cm\hangafter=1
Moran E. C., Filippenko A. V., Chornock R. 2002, ApJ, 579, 71

\noindent\hangindent=0.5cm\hangafter=1
Moretti A., Campana S., Lazzati D., Tagliaferri G. 2003, ApJ, 588, 696

%\noindent\hangindent=0.5cm\hangafter=1
%Mulchaey, J. S., Koraktar, A., Ward, M. J. 1994, ApJ, 436, 586

\noindent\hangindent=0.5cm\hangafter=1
Mulchaey J. S., Davis D. S., Mushotzky R. F., Burstein D. 2003, ApJS, 145, 39

\noindent\hangindent=0.5cm\hangafter=1
Mushotzky R. F., Cowie L. L., Barger A. J., Arnaud K. A. 2000, Nature, 404, 459

%\noindent\hangindent=0.5cm\hangafter=1
%Nakai, N. 1989, PASJ, 41, 1107

\noindent\hangindent=0.5cm\hangafter=1
Nandra K., Pounds K.A. 1994, MNRAS, 268, 405

\noindent\hangindent=0.5cm\hangafter=1
Nicastro F. 2000, ApJ, 530, 65

%\noindent\hangindent=0.5cm\hangafter=1
%Okajima, T., Ebisawa, K., Kawaguchi, T. 2006, ApJ, 652, 105

\noindent\hangindent=0.5cm\hangafter=1
Pietsch W. N., et al. 2001, A\&A, 365, 174

\noindent\hangindent=0.5cm\hangafter=1
Ptak A., Heckman T., Levenson N. A., Weaver K., Strickland D. 2003, ApJ, 592, 782

%\noindent\hangindent=0.5cm\hangafter=1
%Ptak, A., Griffiths, R. 1999, ApJ, 517L, 85P

%\noindent\hangindent=0.5cm\hangafter=1
%Raymond, J. C., Smith, B. W. 1977, ApJS, 35, 419R

\noindent\hangindent=0.5cm\hangafter=1
Read A. M., Stevens I. R. 2002, MNRAS, 335, 36

\noindent\hangindent=0.5cm\hangafter=1
Rigby J. R., Rieke G. H., Donley J. L., Alonso-Herrero A., P\'erez-Gonz\'alez P. G. 2006, ApJ, 645, 115

\noindent\hangindent=0.5cm\hangafter=1
Roberts T. P. 2007, Ap$\&$SS, 311, 203

%\noindent\hangindent=0.5cm\hangafter=1
%Satyapal, S., Dudik, R. P., O'Halloran, B., Gliozzi, M. 2005, ApJ, 663, 86

%\noindent\hangindent=0.5cm\hangafter=1
%Schurch, N. J., Roberts, T. P., Warwick, R. S. 2002, MNRAS, 335, 241S

\noindent\hangindent=0.5cm\hangafter=1
Schlegel D. J., Finkbeiner D. P., Davis M. 1998, ApJ, 500, 525

\noindent\hangindent=0.5cm\hangafter=1
Servillat M., Farrell S. A., Lin D., et al. 2011, arXiv, 1108, 4405

\noindent\hangindent=0.5cm\hangafter=1
Silverman J. D,. et al. 2010, ApJS, 191, 124

\noindent\hangindent=0.5cm\hangafter=1
Smith R., Vaughan S. 2007, MNRAS, 375, 1479

\noindent\hangindent=0.5cm\hangafter=1
Sobolewska M. A., Papadakis I. E. 2009, MNRAS, 399, 1597

%\noindent\hangindent=0.5cm\hangafter=1
%Soria, R., Gosh, K. K. 2009, ApJ, 696, 287S

\noindent\hangindent=0.5cm\hangafter=1
Soria R., Hau G. K., Graham A. W., et al. 2010, MNRAS, 405, 870

%\noindent\hangindent=0.5cm\hangafter=1
%Stobbart, A. M., Roberts, T. P., Wilms, J. 2006, MNRAS, 368, 397

\noindent\hangindent=0.5cm\hangafter=1
Szokoly G. P., et al. 2004, ApJS, 155, 271

\noindent\hangindent=0.5cm\hangafter=1
Taylor R. D., Uttley P., McHardy I. M. 2003, MNRAS, 342, 31

\noindent\hangindent=0.5cm\hangafter=1
Treister E., Urry M. C., Lira P. 2005, ApJ, 630, 104

\noindent\hangindent=0.5cm\hangafter=1
Trouille M., Barger A. J., Cowie L. L., Yang Y., Mushotzky R. F. 2008, ApJS, 179, 1

\noindent\hangindent=0.5cm\hangafter=1
Trump J. R. et al. 2009, ApJ. 706, 797

\noindent\hangindent=0.5cm\hangafter=1
Trump J. R., et al. 2011, arXiv, 1103, 0276

%\noindent\hangindent=0.5cm\hangafter=1
%Tsch\"oke, D., Hensler, G. 2000, RMxAC, 9, 51T

\noindent\hangindent=0.5cm\hangafter=1
Vaughan S., Reeves J., Warwick R., Edelson R. 1999, MNRAS, 309, 113

%\noindent\hangindent=0.5cm\hangafter=1
%Vignatti, P., Molendi, S., Matt, G., et al. 1999, A$\&$A, 349, 57

\noindent\hangindent=0.5cm\hangafter=1
Walton D., Roberts T. P., Mateos S., Heard V. 2011, MNRAS, 416, 1844

\noindent\hangindent=0.5cm\hangafter=1
Wang J--M., Zhang E--P. 2007, ApJ, 660, 1072

\noindent\hangindent=0.5cm\hangafter=1
Warwick R. S., Jenkins L. P., Read A. M., Roberts T. P., Owen R. A. 2007, MNRAS, 376, 1611

\noindent\hangindent=0.5cm\hangafter=1
Watson M. G., et al. 2009, A$\&$A, 493, 339

%\noindent\hangindent=0.5cm\hangafter=1
%Weedman, D. W. 1983, ApJ, 266, 479

\noindent\hangindent=0.5cm\hangafter=1
Wilms J., Allen A., McCray R. 2000, ApJ, 254, 914

\noindent\hangindent=0.5cm\hangafter=1
Worsley M. A., Fabian A. C., Alexander D. M., Bauer F. E., Brandt W. N., Hasinger G., Lehmer B. D. 2005, AIPC, 801, 51 

\noindent\hangindent=0.5cm\hangafter=1
Xue Y. Q., et al. 2010, ApJ, 720, 368 

\noindent\hangindent=0.5cm\hangafter=1
Yuan F., Narayan R. 2004, 612, 724

\noindent\hangindent=0.5cm\hangafter=1
Zezas A., Alonso-Herrero A., Ward M. 2001, Ap$\&$SS, 276, 601 

\end{document}